\documentclass[prd,aps,eqsecnum]{revtex4}
\usepackage{epstopdf}
\usepackage{graphicx}
\usepackage{amsfonts}
\usepackage{rotating}
\usepackage{sparticles}
\usepackage{dcolumn}
\usepackage{bm}
\usepackage{hyperref}
\hypersetup{
    bookmarks=true,         
    unicode=false,          
    pdftoolbar=true,        
    pdfmenubar=true,        
    pdffitwindow=false,     
    pdfstartview={FitH},    
    pdftitle={My title},    
    pdfauthor={Author},     
    pdfsubject={Subject},   
    pdfcreator={Creator},   
    pdfproducer={Producer}, 
    pdfkeywords={keyword1} {key2} {key3}, 
    pdfnewwindow=true,      
    colorlinks=true,       
    linkcolor=red,          
    citecolor=cyan,        
    filecolor=magenta,      
    urlcolor=green,           
    linktocpage=true
}
\usepackage{amsmath,amssymb,amsthm,graphicx,latexsym}
\usepackage{subfigure}
\usepackage{pstricks}
\usepackage{color}

\usepackage{mathrsfs}

\newcommand{\be}{\begin{equation}}
\newcommand{\ee}{\end{equation}}
\newcommand{\bea}{\begin{eqnarray}}
\newcommand{\eea}{\end{eqnarray}}
\newcommand{\bml}{\begin{subequations}}
\newcommand{\eml}{\end{subequations}}
\newcommand{\bfig}{\begin{figure}}
\newcommand{\efig}{\end{figure}}

\begin{document}

\title{Higgs inflation from new K$\ddot{a}$hler potential}

\author{{{Sayantan Choudhury}}\footnote{Electronic address: {sayanphysicsisi@gmail.com}} ${}^{}$
,{{ Trina Chakraborty}}\footnote{Electronic address: {trinachakraborty.27@gmail.com}} ${}^{}$ and {{Supratik Pal}}
\footnote{Electronic address: {supratik@isical.ac.in
}} ${}^{}$}
\affiliation{Physics and Applied Mathematics Unit, Indian Statistical Institute, 203 B.T. Road, Kolkata 700 108, India
}

\date{\today}
\begin{abstract}
We introduce a new class of models of Higgs inflation using the superconformal approach
to supergravity by modifying the K$\ddot{a}$hler geometry. Using the above-mentioned mechanism, we construct a phenomenological functional form of
a new K$\ddot{a}$hler potential followed by construction of various types of models which are characterized by a superconformal
symmetry breaking parameter $\chi$. Depending on the numerical values of $\chi$ we classify the proposed
models into three categories. Models with minimal coupling are identified by $\chi=\pm\frac{2}{3}$ branch
which are made up of shift symmetry preserving flat directions. We also propose various other models
by introducing a non-minimal coupling of the inflaton field to gravity described by $\chi\neq\frac{2}{3}$ branch.
 We employ all these proposed models to study the inflationary paradigm by estimating the major cosmological observables and
confront them with recent observational data from WMAP9 along with other complementary data sets, as well as independently with PLANCK.
 We also mention an allowed range of non-minimal couplings and the {\it Yukawa} type of couplings
appearing in the proposed models used for cosmological parameter estimation.

\end{abstract}


\maketitle
\tableofcontents
\section{\bf Introduction}
\label{intro}

Cosmological inflation has been a paradigm in
which the pathological problems of the Standard Big Bang Cosmology are addressed in a sophisticated way.
The inflaton field yields scale-dependent nearly Gaussian spectrum of density fluctuations.
Moreover, during the inflationary epoch, cosmological perturbation via quantum fluctuation provides seed for the large-scale structure
formation as we perceive today. Inflation is governed by a flat potential which
has a proper field theoretic origin \cite{lyth,lyth2,anupam1,sayan1,sayananu}. In this context, supersymmetry (SUSY) or its local extension (i.e. supergravity (SUGRA))
is the most successful candidate, which imposes certain constraints on the non-supersymmetric models of particle physics and
cosmology \cite{bezru,bezru2,bezru3,simone}. A well known example of such restrictions is the fact that the supersymmetric version of
the Standard Model (SM) of particle physics requires at least two Higgs superfields \cite{nilles,sayan1}. On the other hand,
 the SUSY embedding of the Higgs
model in inflation requires SUGRA \cite{einhorn1,reneta1,reneta2,reneta3,lee1,takahashi,arai2,arai1}. Thus, it is interesting to see how SUSY may affect various inflationary models,
where the gravity sector is minimally or non-minimally coupled to scalar fields \cite{einhorn1,linde1}.

A competent idea is to exercise the SM Higgs doublet as the
inflaton\cite{cota1,cota2} through the well known Higgs inflation{\cite{einhorn2}} within the SUGRA domain\cite{reneta1,reneta2,reneta3,takahashi}.
 In this framework, inflation is realized via a large non-minimal coupling of Higgs doublet to Einstein gravity instead of a
 tiny Higgs quartic coupling, as it contradicts the observed Higgs mass bound at {\it Large Hadron Collider} (LHC) \cite{lhc}.
Earlier it has been shown in various works \cite{burg,barbon,hert,germani} that by applying power counting formalism, Hubble scale during inflation
 approaches the unitarity bound on the new scale in conjunction with the breakdown of the semi-classical approximation
 in the effective field theory of inflation in four dimension below the Ultra-Violet ({\bf UV}) cut-off. However, a customary notion is prevalent
amongst physicists for the study of effective field theory of inflation in which a singlet field with non-minimal coupling can act as a inflaton for a small
singlet self interaction motivated quartic coupling. In such cases Hubble scale can be smaller than the unitarity
bound. The well posed hierarchy problem in the context of SM has been resolved by implementing the well-known weak scale SUSY \cite{nilles,martin}, which is
one of the most important topic of research in particle physics collider phenomenology.
In the framework of Minimal Supersymmetric Standard Model (MSSM) \cite{anupam1,nilles,martin,anupam3}, there is an existence of two Higgs doublets
and the equivalent self coupling can be expressed in terms of the electroweak (EW) gauge couplings.
Setting apart the unitarity problem in the context of MSSM Higgs inflation,
 the implementation of Higgs inflation without fine tuning \cite{arindam} is inconceivable due to the appearance of instability in the ratio of two Higgs VEVs.
 An interesting situation may emerge when the superpotential term provides the vacuum energy via the introduction of an additional self
 interacting coupling
required for inflation governed by Next-to-Minimal Supersymmetric Standard Model (NMSSM) \cite{lyth,reneta1,reneta2,lee1,ulrich}.
 As this new self coupling can be made small without any
violation of the recently observed {\it LHC} bound on the Higgs mass, there might be another physical
possibility appearing where the Higgs inflation can be performed within the semi-classical limit of effective field theory.

However the supergravity theory has a dark side in the context of Higgs inflation.
The main problem
was rooted in the functional form of the K$\ddot{a}$hler potential which involves typical
contributions proportional to quadratic combination of the superfields in the canonical version.
One elegant way to overcome such problem is to search for shift symmetry \cite{reneta3,koushik,reneta4,ido} protected flat directions in supergravity
which can take part in inflation. The flatness of the potential
is broken only by introducing a superconformal symmetry breaking parameter in the supergravity K$\ddot{a}$hler potential \cite{reneta1,reneta2}.
Such terms are directly connected with non-minimal interactions of the inflaton field to the Einstein gravity sector.
This class of non-minimal models of K$\ddot{a}$hler potential has many interesting features, which were explored in the context of superconformal approach to supergravity. Specifically, in the context of canonical superconformal supergravity (CSS)
models \cite{reneta1,reneta2,lee1}, kinetic terms in the preferred frame of reference (Jordan frame) are canonical and the corresponding potential is exactly
same as that appearing in global supersymmetry.
For this purpose, in this article we propose a phenomenological model of a new K$\ddot{a}$hler potential
with two singlet chiral
superfields ($H,S$) which successfully address the problems of supergravity inflation with non-minimal coupling $(\xi_{1},\xi_{2})$.
Here one singlet field plays the role of inflaton and the other one is the background which will trigger preheating \cite{ling,rouz}/reheating
\cite{sayan2,sayan3,anupam4} depending on the branching ratios of different decay channels of the inflaton. Our result can be applied directly to the Higgs inflation by
satisfying D-flat constraints. In this article, our primary target is to do a thorough survey of inflationary models from K$\ddot{a}$hler potential using superconformal transformation
followed by confrontation with latest observational data from {\it WMAP9} \cite{wmap9} and other complementary datasets. The results have also been confronted independently with
PLANCK data \cite{pl1}

The paper is organized as follows. We first explain a general framework for ${\cal N}$=1, ${\cal D}$=4 Jordan
frame supergravity where superconformal symmetry breaking parameters for scalar fields are suitably implemented.
Then we introduce a new phenomenological model of K$\ddot{a}$hler potential with two singlet chiral superfields. Next we discuss the implication of the Higgs
inflation from various types of inflationary potentials derived from the Jordan frame K$\ddot{a}$hler potential for four distinct physical branches
of the symmetry breaking parameter ($\chi$). Next imposing the constraints
from {\it LHC} we employ these models for cosmological parameter estimation by using a numerical code {\it CAMB} \cite{camb}. Finally, we confront the cosmological observables
with the latest available datasets.


\section{\bf Superconformal mechanism in K$\ddot{a}$hler geometry}
\label{model}
In this section we start our discussion with ${\cal N}$=1, ${\cal D}$=4 SUGRA action in the {\it Jordan frame }
 with generalized frame function ${\bf \Phi}(z,\bar{z})$ in the Planckian unit described by \cite{reneta1,reneta2}
\be\label{sug1}
S_{{\bf \Phi}}=\int d^{4}x \sqrt{-g_{J}}\left[R_{(4)}-2\Lambda_{(4)}+e^{-1}_{(4)}{\cal L}^{{\bf \Phi}}_{SUGRA} \right]
\ee
where
\be\begin{array}{llll}\label{lagsugra}
  \displaystyle  e^{-1}_{(4)}{\cal L}^{{\bf \Phi}}_{SUGRA}:=-\frac{{\bf \Phi}(z,\bar{z})}{6}\left[R_{(4)}-\bar{\Psi}_{\mu}R^{\mu}\right] -\frac{1}{6}\left(\partial_{\mu}{\bf \Phi}\right)
\left(\bar{\Psi}^{\alpha}\gamma_{\alpha}\Psi^{\mu}\right)+{\cal L}_{0}+{\cal L}_{\frac{1}{2}}+{\cal L}_{1}+{\cal L}_{m}+{\cal L}_{mix}+{\cal L}_{d}+{\cal L}_{4f}-V_{J}.
   \end{array}\ee
In equation(\ref{lagsugra}) the notations used are: $\Psi_{\mu}\Rightarrow$ gravitino field, $R^{\mu}\Rightarrow$ gravitino kinetic term, ${\cal L}_{0}\Rightarrow$ scalar d.o.f.,
 ${\cal L}_{\frac{1}{2}}\Rightarrow$ fermion d.o.f., ${\cal L}_{1}\Rightarrow$ vector d.o.f., ${\cal L}_{m}\Rightarrow$ fermion mass term,
 ${\cal L}_{mix}\Rightarrow$ mixing term, ${\cal L}_{d}\Rightarrow$ kinetic D term, ${\cal L}_{4f}\Rightarrow$ four fermion term and the SUGRA potential in
 {\it Jordan frame } is
given by \cite{reneta1,reneta2}
\be\begin{array}{llll}\label{jord}
    \displaystyle V_{J}=\frac{{\bf \Phi}^{2}(z,\bar{z})}{9}\left[e^{{\cal K}(z,\bar{z})}\left\{\left(\nabla_{\alpha}{\cal W}(z)\right)G^{\alpha\bar{\beta}}
\left(\nabla_{\bar{\beta}}{\cal \bar{W}}(z)\right)-3|{\cal W}(z)|^{2}\right\}+\frac{1}{2}\left({\bf Re}~f(z)\right)^{-1~AB}{\cal P}_{A}{\cal P}_{B} \right]
   \end{array}\ee
where $\alpha=1,2,....,n$ represents the number of complex scalars in the SUGRA chiral multiplet, ${\cal K}(z,\bar{z})$ is the K$\ddot{a}$hler potential, ${\cal W}(z)$ is the holomorphic superpotential,
$f_{AB}(z)$ is the holomorphic kinetic gauge matrix field and the Killing potential or momentum map is denoted by ${\cal P}_{A}$
which includes all the Yang-Mills transformation of the scalars through which {\it Fayet-Iliopoulos} terms are also taken care of. In equation(\ref{sug1})
the supergravity {\it verbien} (inverse of {\it f$\ddot{u}$nfbien}) is characterized by the transformation rule \cite{sayan4}
\be\begin{array}{llll}\label{opbvc}
\displaystyle g^{J}_{\mu\nu}:=\eta_{\hat{A}\hat{B}}\left(V^{\hat{A}}_{\mu}\otimes V^{\hat{B}}_{\nu}\right)
\end{array}\ee
with \be\begin{array}{llll}\label{opbvc1}\displaystyle Det(V)=\sqrt{-g_{J}}=e_{(4)}.\end{array}\ee Here
we use the following defintion of covariant derivative:
\be\begin{array}{llll}\label{vb1}
    \displaystyle \nabla_{\alpha}{\cal W}:={\cal W}_{\alpha}+{\cal K}_{\alpha}{\cal W}
   \end{array}\ee
where the subscript $\alpha$ denotes differentiation with respect to complex field $z^{\alpha}$.
By setting ${\bf \Phi}=-3$, the SUGRA action in {\it Jordan frame } reduces to the
well known action in the {\it Einstein frame}. Consequently the potential stated in equation(\ref{jord}) can be related to its
{\it Einstein frame} counterpart as
\be\label{efcount}
V_{J}=\frac{{\bf \Phi}^{2}(z,\bar{z})}{9}V_{E},\ee

where the subscripts $J$ and $E$ are used to denote {\it Jordan } and {\it Einstein} frame. Here both the
 frames are connected via the {\it superconformal transformation} defined in terms of the metric as
\be\label{cft1}
g^{J}_{\mu\nu}=\Omega^{2}(z,\bar{z})g^{E}_{\mu\nu}\ee
where we identify the conformal factor with
\be\label{cf12}\Omega^{2}(z,\bar{z})=-\frac{{\bf \Phi}(z,\bar{z})}{3}=e^{-\frac{{\cal K}(z,\bar{z})}{3}}\ee
which yields a purely bosonic action in ${\cal N}$=1, ${\cal D}$=4 SUGRA in a specific {\it Jordan frame }
 triggering the {\it superHiggs mechanism}.  The SUGRA action includes ${\cal SU}({\bf 2,2|1})$ superconformal
symmetry, local dilation, special conformal symmetry, special SUSY and local ${\cal U}(1)_{R}$ symmetry and other local symmetries of ${\cal N}$=1, ${\cal D}$=4 SUGRA.
Such a superconformal mechanism is very useful to embed a class of scale invariant Global Supersymmetric (GSUSY) models into SUGRA theory. By ``embedding'',
here we actually point towards the fact that the ${\cal N}$=1, ${\cal D}$=4 self-interacting SUGRA multiplets has a local Poincare SUSY which can be obtained by the breakdown of above mentioned
superconformal symmetry. Consequently the pure SUGRA sector in the action stated by equation(\ref{sug1}) breaks superconformal symmetry and the matter part remains
superconformal after gauge fixing. The non-canonical nature of the kinetic term is generally guaranteed by the following choice of {\it superconformal factor} \cite{reneta1,reneta2,reneta3}:
\be\begin{array}{llll}\label{ar1}
    \displaystyle \Omega^{2}(z,\bar{z})=1-\frac{1}{3}\left(\delta_{\alpha\bar{\beta}}z^{\alpha}\bar{z}^{\bar{\beta}}+{\cal J}(z)+\bar{\cal J}(\bar{z})\right)
   \end{array}\ee
where ${\cal J}(z)$ and $\bar{\cal J}(\bar{z})$ are the phenomenological holomorphic functions considered in the K$\ddot{a}$hler gauge.
It is important to mention here that the dilation symmetry implies $\Omega^{2}(z,\bar{z})$ to be homogeneous of first degree in both $z$ and $\bar{z}$,
${\cal W}(z)$ to be homogeneous of third degree in $z$. Additionally local ${\cal U}(1)_{R}$ symmetry implies $\Omega^{2}(z,\bar{z})$ is neutral and
${\cal W}(z)$ has chiral weight three (which has been taken care of in equation(\ref{cz1})). We also assume that the resultant potential is obtained only
from the supergravity F-term as the kinetic sector is gauge fixed by imposing the D-flat constraints.

Now using equation(\ref{cf12}) and
equation(\ref{ar1}) one can find out the explicit expressions for SUGRA {\it frame function} and {\it K$\ddot{a}$hler potential} in this context. Using these
results we obtain the following expression for {\it K$\ddot{a}$hler} metric:
\be\begin{array}{lllll}\label{klqw}
    \displaystyle G^{\alpha\bar{\beta}}=\left(\frac{\partial^{2}\Omega^{2}(z,\bar{z})}{\partial z_{\alpha}\partial \bar{z}_{\bar{\beta}}}\right)=\left\{1-\frac{1}{3}\left(\delta_{\alpha\bar{\beta}}z^{\alpha}\bar{z}^{\bar{\beta}}+{\cal J}(z)+\bar{\cal J}(\bar{z})\right)
\right\}\left[\delta^{\alpha\bar{\beta}}-\frac{1}{3}\left(z^{\alpha}\bar{z}^{\bar{\beta}}
+\delta^{\alpha\bar{\beta}}\left({\cal J}(z)+\bar{\cal J}(\bar{z})\right)\right)\right]
   \end{array}\ee

Assuming non-canonical structure of the {\it superconformal factor} stated in equation(\ref{ar1})
 let us prove the equivalence of F-term SUGRA potential in superconformal {\it Jordan frame } and in GSUSY.
We start with a renormalizable ${\cal N}$=1, ${\cal D}$=4 SUGRA where
the most generalized expression of the superpotential is constrained to the following cubic form:
\be\begin{array}{llll}\label{cz1}
    \displaystyle {\cal W}(z)=\frac{1}{3}{\bf d}_{\alpha\beta\gamma}z^{\alpha}z^{\beta}z^{\gamma}
   \end{array}\ee
where ${\bf d}_{\alpha\beta\gamma}$'s are the trilinear couplings in SUGRA theory.
Equation(\ref{cz1}) breaks the ${\bf {\cal SU}(1,n)}$ symmetry.
Now considering the fact that the SUGRA superpotential
is homogeneous of the third degree in $z^{\alpha}$'s we get:
 \be\begin{array}{llll}\label{conaz}
\displaystyle {\cal W}_{\alpha}z^{\alpha}=3{\cal W},\\ \displaystyle \bar{\cal W}_{\bar{\alpha}}\bar{z}^{\bar{\alpha}}=3\bar{\cal W}.
\end{array}\ee
Considering all the above facts the {\it Jordan frame } D-flat potential turns out to be
\be\begin{array}{llll}\label{zxcq1}
    \displaystyle V^{F}_{J}=\left(1-\frac{1}{3}\left({\cal J}(z)+\bar{\cal J}(\bar{z})\right)\right)\left[V^{F}_{GSUSY}(z)
+\bar{\cal W}\left(\partial_{z^{\alpha}}{\cal J}(z)\right)+\bar{\cal W}\left(\partial_{\bar{z}^{\bar{\alpha}}}\bar{\cal J}(\bar{z})\right)\right]+|{\cal W}|^{2}
\left\{\delta_{\alpha\bar{\beta}}z^{\alpha}\bar{z}^{\bar{\beta}}+{\cal J}(z)+\bar{\cal J}(\bar{z})\right.\\
\left.~~~~~~~\displaystyle \left(1-\frac{1}{3}\left({\cal J}(z)+\bar{\cal J}(\bar{z})\right)\right)
\left[\delta_{\bar{\gamma}\lambda}\bar{z}^{\bar{\gamma}}z^{\lambda}+z^{\alpha}\left(\partial_{z^{\alpha}}{\cal J}(z)\right)
+\bar{z}^{\bar{\beta}}\left(\partial_{\bar{z}^{\bar{\alpha}}}\bar{\cal J}(\bar{z})\right)+\delta^{\alpha\bar{\beta}}\left(\partial_{z^{\alpha}}{\cal J}(z)\right)
\left(\partial_{\bar{z}^{\bar{\alpha}}}\bar{\cal J}(\bar{z})\right)\right]
\right.\\ \left. ~~~~~~~
\displaystyle -\frac{1}{3}z^{\alpha}\bar{z}^{\bar{\beta}}\left[\delta_{\alpha\bar{\gamma}}\delta_{\bar{\beta}\alpha^{'}}\bar{z}^{\bar{\gamma}}z^{\alpha^{'}}
+\delta_{\bar{\beta}\alpha^{'}}z^{\alpha^{'}}\left(\partial_{z^{\alpha}}{\cal J}(z)\right)+\delta_{\alpha\bar{\gamma}}\bar{z}^{\bar{\gamma}}
\left(\partial_{\bar{z}^{\bar{\beta}}}\bar{\cal J}(\bar{z})\right)+\left(\partial_{z^{\alpha}}{\cal J}(z)\right)\left(\partial_{\bar{z}^{\bar{\beta}}}\bar{\cal J}(\bar{z})\right)
\right]\right\}\\
\displaystyle~~~~~~~-\frac{1}{3}z^{\alpha}\bar{z}^{\bar{\beta}}\left\{3|{\cal W}|^{2}\delta_{\alpha\bar{\beta}}
+{\cal W}\bar{\cal W}_{\bar{\beta}}\left(\partial_{z^{\alpha}}{\cal J}(z)\right)+\delta_{\bar{\beta}\gamma}\bar{\cal W}{\cal W}_{\alpha}z^{\gamma}
+{\cal W}{\cal W}_{\alpha}\left(\partial_{\bar{z}^{\bar{\beta}}}\bar{\cal J}(\bar{z})\right)\right\}
   \end{array}\ee

where GSUSY potential $V_{GSUSY}(z)=\delta^{\alpha\bar{\beta}}{\cal W}_{\alpha}\bar{\cal W}_{\bar{\beta}}$. Here the superscript $F$ denotes F-term potential.
Here it is important to mention that when {\it superconformal symmetry is gauge fixed,}
the matter multiplets are preserved, which implies  ${\cal J}(z)=0$ and $\bar{\cal J}(\bar{z})=0$.
Consequently equation(\ref{zxcq1}) reduces to the following D-flat form of the effective potential:
\be\begin{array}{llll}\label{gsusy}
    \displaystyle V^{F}_{J}=V^{F}_{GSUSY}(z)-\frac{1}{3}\delta_{\alpha\bar{\gamma}}
\delta_{\bar{\beta}\alpha^{'}}z^{\alpha}\bar{z}^{\bar{\beta}}\bar{z}^{\bar{\gamma}}z^{\alpha^{'}}|{\cal W}|^{2}
   \end{array}\ee
where in the last non-renormalizable term of the above expansion the superpotential is highly suppressed by the {\bf UV} cut-off scale ($\Lambda_{UV}$) of the effective theory
of gravity in presence of ${\cal O}(1/\Lambda^{2}_{UV})$ order term. Here $\Lambda_{UV}$ is fixed at the value of reduced Planck scale 
$M_{PL}(\sim ~2.43\times10^{18}{\rm GeV})$ in the Planckian unit system beyond which the theory becomes unprotective from {\bf UV} end 
and the effective field theory prescription doesn't hold good
in our proposed setup. The contribution from the last term of Eq~(\ref{gsusy}) originates from the
 quadratically Planck scale suppressed higher dimensional K\"ahler operators in ${\cal N}=1$ 
SUGRA theory. Most importantly, in four dimension, such K\"ahler corrections doesn't contribute to the leading order of effective theory.
 Consequently below such high scale {\bf UV} cut-off, renormalizability of the effective potential is automatically demanded within the effective theory prescription 
and
finally we have:
\be\label{absent}
V^{F}_{J}\simeq V^{F}_{GSUSY}(z\leq \Lambda_{UV}=M_{Pl})
\ee
leading to the equivalence of F-term potentials as claimed above.
Next we will concentrate on a specific situation where the superconformal symmetry is broken via the non-minimal coupling parameter $\chi$ with
gravity. Consequently the frame function stated in equation(\ref{ar1}) is modified as \cite{reneta1,reneta2}:
\be\begin{array}{llll}\label{modi}
    \displaystyle \Omega^{2}(z,\bar{z})=-|z^{0}|^{2}+|z^{\alpha}|^{2}-\chi\left(\Theta_{\alpha\beta}\frac{z^{\alpha}z^{\beta}\bar{z}^{\bar{0}}}{z^{0}}+
\bar{\Theta}_{\alpha\beta}\frac{\bar{z}^{\bar{\alpha}}\bar{z}^{\bar{\beta}}z^{0}}{\bar{z}^{\bar{0}}}\right)
   \end{array}\ee
which characterizes the non-flat moduli space geometry in SUGRA. Now gauge fixing criteria demands that in Planckian Unit system the compensator fields satisfy
$z^{0}=\bar{z}^{\bar{0}}=\sqrt{3}$.
This implies a subsequent modification in the matter part of the inverse K$\ddot{a}$hler metric of the enlarged space which can be expressed as:
\be\begin{array}{llll}\label{kler}
    \displaystyle G^{\alpha\bar{\beta}}=\delta^{\alpha\bar{\beta}}-\frac{4\chi^{2}\delta^{\alpha\bar{\lambda}}\delta^{\sigma\bar{\beta}}\Theta_{\sigma\zeta}
\bar{\Theta}_{\bar{\lambda}\bar{\xi}}z^{\zeta}\bar{z}^{\bar{\xi}}}{\left[3-\chi\left(\Theta_{\gamma\eta}z^{\gamma}z^{\eta}
+\bar{\Theta}_{\bar{\gamma}\bar{\eta}}\bar{z}^{\bar{\gamma}}\bar{z}^{\bar{\eta}}\right)+4\chi^{2}\delta^{\gamma\bar{\eta}}\Theta_{\gamma\zeta}
\bar{\Theta}_{\bar{\eta}\bar{\rho}}z^{\zeta}\bar{z}^{\bar{\rho}}\right]},\\
\displaystyle G^{0\bar{\beta}}=-\frac{2\sqrt{3}\chi\delta^{\lambda\bar{\beta}}\Theta_{\lambda\xi}z^{\xi}}{\left[3
-\chi\left(\Theta_{\gamma\eta}z^{\gamma}z^{\eta}+\bar{\Theta}_{\bar{\gamma}\bar{\eta}}\bar{z}^{\bar{\gamma}}\bar{z}^{\bar{\eta}}\right)
+4\chi^{2}\delta^{\gamma\bar{\eta}}\Theta_{\gamma\rho}\bar{\Theta}_{\bar{\eta}\bar{\sigma}}z^{\rho}\bar{z}^{\bar{\sigma}}\right]},\\
\displaystyle G^{\alpha\bar{0}}=-\frac{2\sqrt{3}\chi\delta^{\alpha\bar{\lambda}}\bar{\Theta}_{\bar{\lambda}\bar{\xi}}\bar{z}^{\bar{\xi}}}{\left[3
-\chi\left(\Theta_{\gamma\eta}z^{\gamma}z^{\eta}+\bar{\Theta}_{\bar{\gamma}\bar{\eta}}\bar{z}^{\bar{\gamma}}\bar{z}^{\bar{\eta}}\right)
+4\chi^{2}\delta^{\gamma\bar{\eta}}\Theta_{\gamma\rho}\bar{\Theta}_{\bar{\eta}\bar{\sigma}}z^{\rho}\bar{z}^{\bar{\sigma}}\right]},\\
\displaystyle G^{0\bar{0}}=-\frac{3}{\left[3-\chi\left(\Theta_{\gamma\eta}z^{\gamma}z^{\eta}
+\bar{\Theta}_{\bar{\gamma}\bar{\eta}}\bar{z}^{\bar{\gamma}}\bar{z}^{\bar{\eta}}\right)+4\chi^{2}\delta^{\gamma\bar{\eta}}\Theta_{\gamma\zeta}
\bar{\Theta}_{\bar{\eta}\bar{\rho}}z^{\zeta}\bar{z}^{\bar{\rho}}\right]},
   \end{array}\ee
subject to the {\it orthonormalization condition}
\be\label{jkq1}G^{0\bar{\beta}}G_{0\bar{\gamma}}+G^{\alpha\bar{\beta}}G_{\alpha\bar{\gamma}}=\delta^{\bar{\beta}}_{\bar{\gamma}}.\ee
This will directly modify the {\it Jordan frame} potential stated in equation(\ref{jord}). In the next two sections we will discuss elaborately
the cosmological consequences of such non-minimal coupling parameter in the context of {\it superHiggs} theory.


\section{\bf Inflationary model building for different values of the non-minimal coupling ($\chi$)}
\label{dreduc}

\begin{figure}[htb]
{\centerline{\includegraphics[width=13cm, height=8cm] {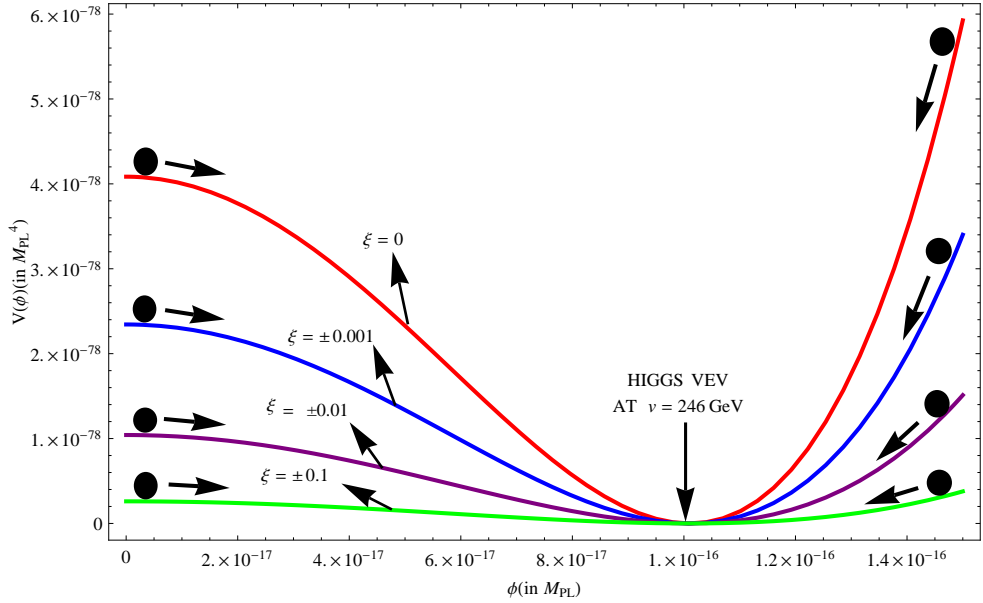}}}
\caption{Family of Higgs potentials for
different numerical values of non-minimal coupling $\xi$, starting from $\xi = 0$. Inflation occurs either when the field $\phi=(H,S)$ rolls
down from its large numerical values or when it rolls down from $\phi = 0$.
  } \label{figpot}
\end{figure}

In this section we will start our discussion with a simple gauge fixed version of frame function
in the presence of a superconformal symmetry breaking term ($\chi$) in the Planckian unit:
\be\begin{array}{llll}\label{vb1}
    \displaystyle {\bf \Phi}(H,S,\bar{H},\bar{S})=-3-\frac{1}{4}\left(1+\frac{3\chi}{2}\right)\left[(H-\bar{H})^{2}+(S-\bar{S})^{2}\right]
+\frac{1}{4}\left(1-\frac{3\chi}{2}\right)\left[(H+\bar{H})^{2}+(S+\bar{S})^{2}\right].
   \end{array}\ee
Using equation(\ref{cf12}), the conformal factor turns out to be:
\be\begin{array}{llll}\label{cvt1}
    \displaystyle \Omega^{2}(H,S,\bar{H},\bar{S})=1+\frac{1}{12}\left(1+\frac{3\chi}{2}\right)\left[(H-\bar{H})^{2}+(S-\bar{S})^{2}\right]
-\frac{1}{12}\left(1-\frac{3\chi}{2}\right)\left[(H+\bar{H})^{2}+(S+\bar{S})^{2}\right].
   \end{array}\ee
Here the superHiggs sector $H=\frac{H_{1}+iH_{2}}{\sqrt{2}}$ and $S=\frac{S_{1}+iS_{2}}{\sqrt{2}}$ are complex scalar fields in the SUGRA chiral multiplet.
Depending on the numerical values of $\chi$, shift symmetry of H and S fields are preserved, which is one of the necessary tools to resolve {\it SUGRA $\eta$~problem} in
the context of inflation. However, the model will suffer from the well known {\it tachyonic mass problem} \cite{reneta1,reneta2,lee1,sen1} in superHiggs theory, which can be resolved
by adding higher order non-minimal quartic correction terms $\beta_{1} (H\bar{H})^2$ or $\beta_{2} (S\bar{S})^2$
in the frame function as well as in the conformal factor stated in equation(\ref{vb1}) and equation(\ref{cvt1}) respectively. Here $(\beta_{1},\beta_{2})$ are two dimensional
non-minimal couplings which are highly suppressed by the {\bf UV} cut-off scale of the effective theory by ${\cal O}(1/M^{4}_{PL})$ order term 
in Planckian unit. Once we add such corrections to the proposed model, {\it tachyonic mass problem} is resolved immediately in the next to leading
 order of the effective theory. But
this will explicitly break the shift symmetry, the result of which is reappearance of {\it SUGRA $\eta$~problem}. 
However, in our prescribed effective field theory setup the {\it tachyonic mass problem} will not at all appear as the 
VEV of the Higss field is too small compared to the {\bf UV} cut-off of the effective theory ($\Lambda_{UV}=M_{PL})$ and the scale of superHiggs inflation
($\sqrt[4]{V_{inf}}\sim 4.11\times 10^{-3}_{PL}\sim M_{GUT})$. Here we fix the VEV of the Higgs field at $v=1.01\times10^{-16}M_{PL}\sim 246~{\rm GeV}$,
 which sets the Higgs mass at the observed value $m_{H}=5.14\times10^{-17}M_{PL}\sim 125~{\rm GeV}$ by LHC \cite{lhc}.
The behavior of the Higgs potential for various values of
the non-minimal coupling is explicitly shown in figure(\ref{figpot}). This shows that with the increasing strengths of non-minimal coupling, the
corresponding potential becomes more and more flat. In
the next subsections we will study the cosmological consequences of these models in detail.


\subsection{\bf Models with $\chi=\frac{2}{3}$}
In this branch the conformal factor is given by:
\be\begin{array}{llll}\label{cvt1x}
    \displaystyle \Omega^{2}(H,S,\bar{H},\bar{S})=1+\frac{1}{6}\left[(H-\bar{H})^{2}+(S-\bar{S})^{2}\right]
   \end{array}\ee
which is connected to the K$\ddot{a}$hler potential via equation(\ref{cf12}). In this context the following transformations
\be\begin{array}{llll}\label{opi}
    \displaystyle H\rightarrow H+C_{H},\\
\displaystyle ~S\rightarrow S+C_{S}
   \end{array}\ee
lead to the shift symmetry of the K$\ddot{a}$hler potential with respect to $(H-\bar{H})$ and $(S-\bar{S})$,
 provided $C_{H}$ and $C_{S}$ are constant shifts along real axis of H and S complex plane.

\begin{table}[htb]\small\addtolength{\tabcolsep}{-1pt}
   \small\begin{tabular}{|c|c|c|c|c|}
   \hline ${\bf Class~of~models}$  &{\bf $\Omega^{2}$ }
&  ${\bf {\cal W}}$ &{\bf $V_{J}$}
&  {\bf $V_{E}$}  \\
\hline H~real,S=0 & 1 & -$\lambda_{1} S\left(H\bar{H}-\frac{v^{2}_{1}}{2}\right)$
 &$\frac{\lambda^{2}_{1}}{4}\left(H^{2}_{1}-v^{2}_{1}\right)^{2}$ & $\frac{\lambda^{2}_{1}}{4}\left(H^{2}_{1}-v^{2}_{1}\right)^{2}$\\
   \hline
 H=0,S~real  &1 & -$\lambda_{2} H\left(S\bar{S}-\frac{v^{2}_{2}}{2}\right)$ &$\frac{\lambda^{2}_{2}}{4}\left(S^{2}_{1}-v^{2}_{2}\right)^{2}$
 & $\frac{\lambda^{2}_{2}}{4}\left(S^{2}_{1}-v^{2}_{2}\right)^{2}$ \\
    \hline
 H~complex,S=0  & $\left(1-\frac{H^{2}_{2}}{3}\right)$ & -$\lambda_{1} S\left(H\bar{H}-\frac{v^{2}_{1}}{2}\right)$
&$\frac{\lambda^{2}_{1}}{4}\left(H^{2}_{1}+H^{2}_{2}-v^{2}_{1}\right)^{2}$ &$\frac{\frac{\lambda^{2}_{1}}{4}\left(H^{2}_{1}+H^{2}_{2}-v^{2}_{1}\right)^{2}}{\left(1-\frac{H^{2}_{2}}{3}\right)^{2}}$ \\
 \hline H=0,S~complex& $\left(1-\frac{S^{2}_{2}}{3}\right)$&-$\lambda_{2} H\left(S\bar{S}-\frac{v^{2}_{2}}{2}\right)$ & $\frac{\lambda^{2}_{2}}{4}\left(S^{2}_{1}+S^{2}_{2}-v^{2}_{2}\right)^{2}$
&$\frac{\frac{\lambda^{2}_{2}}{4}\left(S^{2}_{1}+S^{2}_{2}-v^{2}_{2}\right)^{2}}{\left(1-\frac{S^{2}_{2}}{3}\right)^{2}}$ \\
   \hline
\end{tabular}
  \caption{\it Jordan frame and Einstein frame potentials
obtained from $\chi=\frac{2}{3}$ branch.}
  \label{tab1}
  \end{table}
\begin{table}[htb]\small\addtolength{\tabcolsep}{-1pt}
\small\begin{tabular}{|l|l|l|l|l|l|l|l|l|l|l|l|l|l|l|}
  \hline
  {\bf Potential} & {\bf Confronts }& {\bf Coup} &{ \bf $P_{S}$}&{\bf $n_{S}$}& {\bf $\alpha_{S}$}& {\bf $r$ }&
 {\bf $\Omega_{\Lambda}$}&{\bf $\Omega_{m}$}&{\bf $\sigma_{8}$}&{\bf $\eta_{Rec}$} &{\bf $\eta_{0}$}\\
 &{\bf with}&{\bf-ling($\lambda_{1}$)} &{\bf ($\times 10^{-9}$)} & &{\bf ($\times 10^{-4}$)} &   & &  & &{\bf Mpc} &{\bf Mpc} \\
 & &{\bf($\times 10^{-7}$)}  &  & & & & & & & &    \\
\hline
   $\frac{\lambda^{2}_{1}}{4}\left({\bf{H^{2}_{1}}}-v^{2}_{1}\right)^{2}$ &${\bf{ \Lambda CDM(WMAP9)/PLANCK}}$ & $1.43$ & $2.354$ & $0.0.958$&
$-5.894 $&$0.048$  & $0.684$&$0.316$ &$0.819 $ &$280.38$ & $14184.8$ \\
  \hline
  $\frac{\frac{\lambda^{2}_{1}}{4}\left({\bf{H^{2}_{1}}}+H^{2}_{2}-v^{2}_{1}\right)^{2}}
{\left(1-\frac{H^{2}_{2}}{3}\right)^{2}}$&${\bf{\Lambda CDM(WMAP9+spt}}$ & $1.250$& $2.321$ & $0.964$&$-4.422 $&$0.046$ &
 $0.684$ & $0.316$ & $0.822$ & $280.38$ &$14184.8$ \\
  &${\bf{ +act+h_{0})/PLANCK}}$ &   & & & &  & & & & &   \\
    \hline
\end{tabular}
\caption{\it Cosmological parameter estimation for observationally allowed models
obtained from $\chi=\frac{2}{3}$ branch. 
}
\label{tab2}
\end{table}
In table(\ref{tab1}) we have listed several classes of {\it Jordan frame} and {\it Einstein frame}
 potentials obtained from all possible physical combinations of H and S of the superconformal transformation mentioned in equation(\ref{cvt1x}).
 In this article, potentials obtained from
$H$ and $S$ in any branch are exactly similar. So we will restrict ourselves to the $H$ dependent models for cosmological parameter estimation.
In order to confront with the {\it recently observed Higgs at
LHC, here we fix the VEV, $v_{1}=246~ GeV$ with mass $125~ GeV$}.

\begin{figure}[htb]
\centering
\subfigure[$TT$]{
    \includegraphics[width=5.5cm,height=4.7cm] {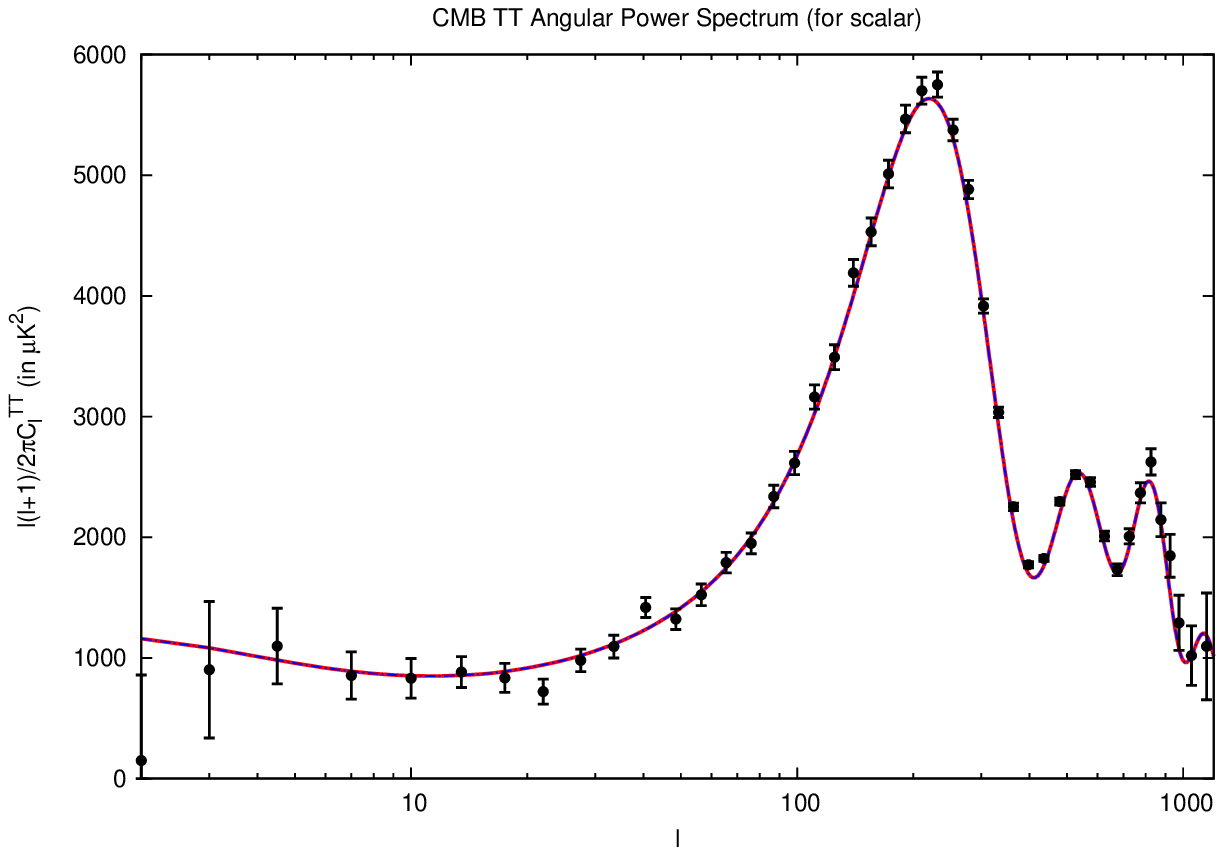}
    \label{fig:subfig1av}
}
\subfigure[$TE$]{
    \includegraphics[width=5.5cm,height=4.7cm] {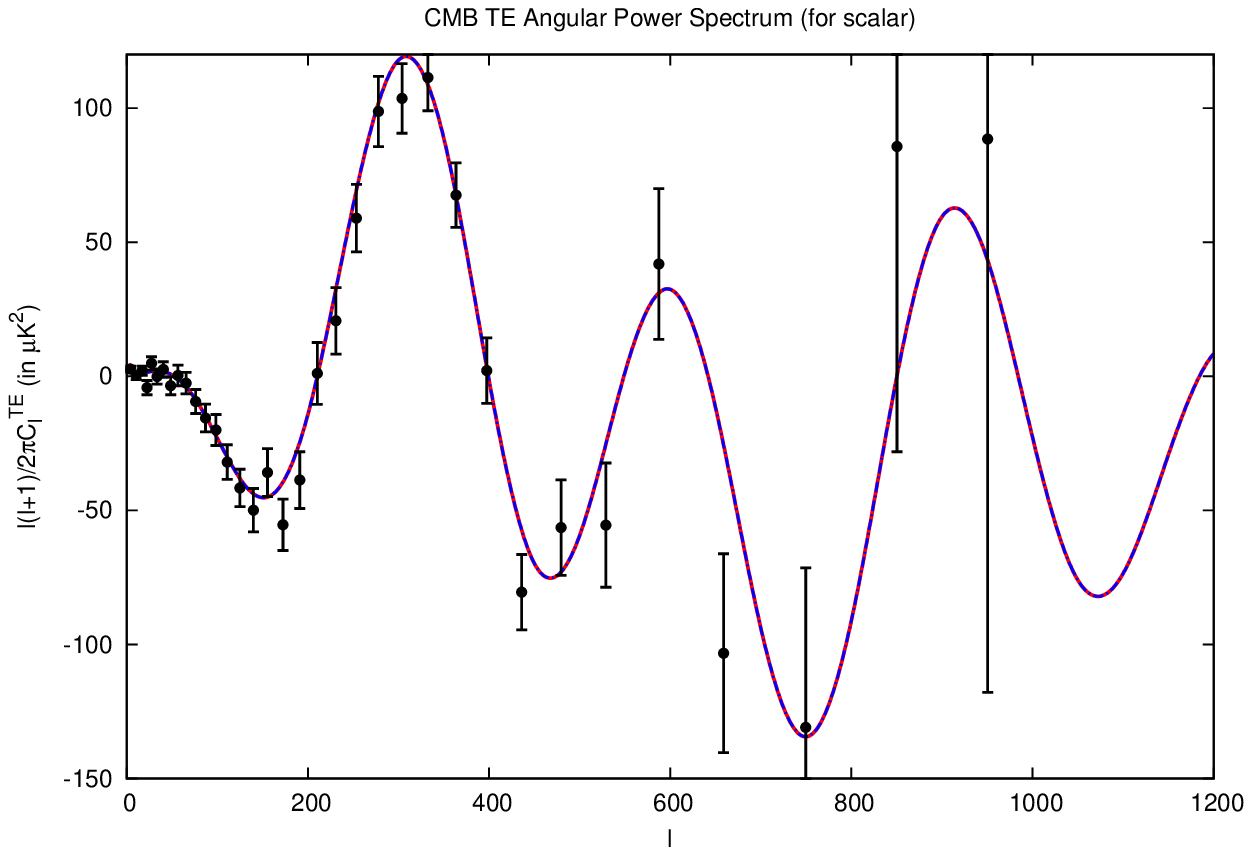}
    \label{fig:subfig2av}
}
\subfigure[$EE$]{
    \includegraphics[width=5.5cm,height=4.7cm] {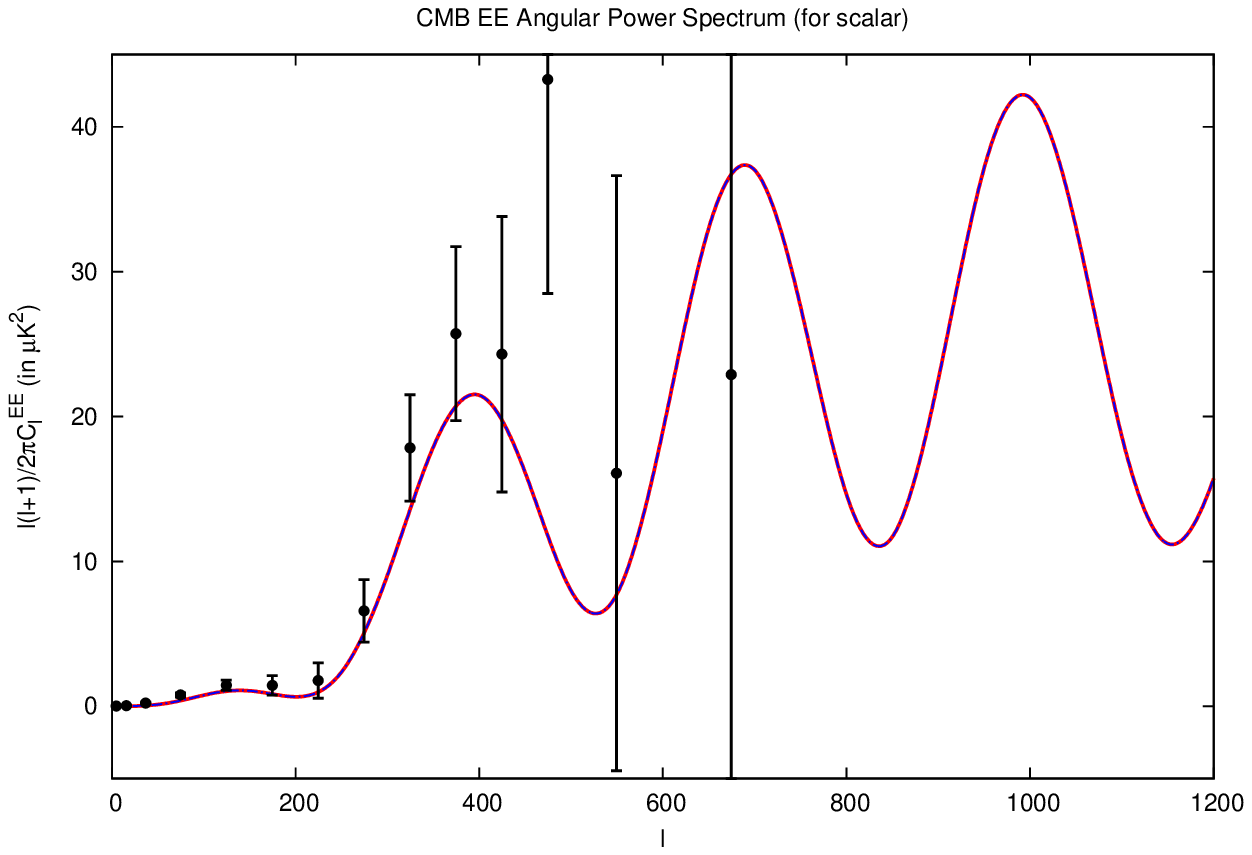}
    \label{fig:subfig3av}
}
\caption[Optional caption for list of figures]{Variation
 of CMB angular power spectra vs multipoles ($l$) for \subref{fig:subfig1av} TT, \subref{fig:subfig2av} TE
and \subref{fig:subfig3av}EE mode from $\chi=\frac{2}{3}$ branch. The statistical error bars are obtained from WMAP9 data}
\label{fig1}
\end{figure}

\subsection{\bf Models with $\chi=-\frac{2}{3}$}
In this branch the conformal factor in equation(\ref{cvt1}) reduces to the following:
\be\begin{array}{llll}\label{cvt1y}
    \displaystyle \Omega^{2}(H,S,\bar{H},\bar{S})=1-\frac{1}{6}\left[(H+\bar{H})^{2}+(S+\bar{S})^{2}\right]
   \end{array}\ee
which is connected to the K$\ddot{a}$hler potential via equation(\ref{cf12}). In this context the following transformations
\be\begin{array}{llll}\label{opi}
    \displaystyle H\rightarrow H+\tilde{C}_{H},\\
\displaystyle ~S\rightarrow S+\tilde{C}_{S}
   \end{array}\ee
lead to the shift symmetry of the K$\ddot{a}$hler potential with respect to $(H+\bar{H})$ and $(S+\bar{S})$,
 provided $\tilde{C}_{H}$ and $\tilde{C}_{S}$ are constant shifts along imaginary axis of H and S complex plane.

\begin{table}[htb]\small\addtolength{\tabcolsep}{-1pt}
   \small\begin{tabular}{|c|c|c|c|c|}
   \hline ${\bf Class~of~models}$  &{\bf $\Omega^{2}$ }
&  ${\bf {\cal W}}$ &{\bf $V_{J}$}
&  {\bf $V_{E}$}  \\
\hline H~real,S=0 & $\left(1-\frac{H^{2}_{1}}{3}\right)$ & -$\lambda_{1} S\left(H\bar{H}-\frac{v^{2}_{1}}{2}\right)$
 &$\frac{\lambda^{2}_{1}}{4}\left(H^{2}_{1}-v^{2}_{1}\right)^{2}$ & $\frac{\frac{\lambda^{2}_{1}}{4}\left(H^{2}_{1}-v^{2}_{1}\right)^{2}}
{\left(1-\frac{H^{2}_{1}}{3}\right)^{2}}$\\
   \hline
 H=0,S~real  &$\left(1-\frac{S^{2}_{1}}{3}\right)$ & -$\lambda_{2} H\left(S\bar{S}-\frac{v^{2}_{2}}{2}\right)$ &$\frac{\lambda^{2}_{2}}{4}\left(S^{2}_{1}-v^{2}_{2}\right)^{2}$
 & $\frac{\frac{\lambda^{2}_{2}}{4}\left(S^{2}_{1}-v^{2}_{2}\right)^{2}}{\left(1-\frac{S^{2}_{1}}{3}\right)^{2}}$ \\
    \hline
 H~complex,S=0  & $\left(1-\frac{H^{2}_{1}}{3}\right)$ & -$\lambda_{1} S\left(H\bar{H}-\frac{v^{2}_{1}}{2}\right)$
&$\frac{\lambda^{2}_{1}}{4}\left(H^{2}_{1}+H^{2}_{2}-v^{2}_{1}\right)^{2}$ &$\frac{\frac{\lambda^{2}_{1}}{4}\left(H^{2}_{1}+H^{2}_{2}-v^{2}_{1}\right)^{2}}{\left(1-\frac{H^{2}_{1}}{3}\right)^{2}}$ \\
 \hline H=0,S~complex& $\left(1-\frac{S^{2}_{1}}{3}\right)$&-$\lambda_{2} H\left(S\bar{S}-\frac{v^{2}_{2}}{2}\right)$ & $\frac{\lambda^{2}_{2}}{4}\left(S^{2}_{1}+S^{2}_{2}-v^{2}_{2}\right)^{2}$
&$\frac{\frac{\lambda^{2}_{2}}{4}\left(S^{2}_{1}+S^{2}_{2}-v^{2}_{2}\right)^{2}}{\left(1-\frac{S^{2}_{1}}{3}\right)^{2}}$ \\
   \hline
\end{tabular}
  \caption{\it Jordan frame and Einstein frame potentials
obtained from $\chi=-\frac{2}{3}$ branch.}
  \label{tab3}
  \end{table}

\begin{table}[htb]\small\addtolength{\tabcolsep}{-1pt}
\small\begin{tabular}{|l|l|l|l|l|l|l|l|l|l|l|l|l|}
  \hline
  {\bf Potential}& {\bf Confronts } & {\bf Coup} &{ \bf $P_{R}$}&{\bf $n_{s}$}& {\bf $\alpha_{s}$}& {\bf $r$ }&
 {\bf $\Omega_{\Lambda}$}&{\bf $\Omega_{m}$}&{\bf $\sigma_{8}$} &{\bf $\eta_{Rec}$} &{\bf $\eta_{0}$}\\
 &{\bf with}&{\bf-ling($\lambda_{1}$)} &{\bf ($\times 10^{-9}$)} & &{\bf ($\times 10^{-4}$)} &  & &  & &{\bf Mpc} &{\bf Mpc} \\
& &{\bf($\times 10^{-7}$)}  &  & & & & & & & &   \\
\hline
  $\frac{\frac{\lambda^{2}_{1}}{4}\left({H^{2}_{1}}+{\bf{H^{2}_{2}}}-v^{2}_{1}\right)^{2}}
{\left(1-\frac{H^{2}_{1}}{3}\right)^{2}}$ &${\bf{ \Lambda CDM(WMAP9+spt}}$&
 $1.250$&$2.321$ &$0.964$ &-$4.422$ & $0.046$ & $0.684$ & $0.316$ & $0.822$ & $280.38$ & $14184.8$ \\
   &${\bf{ +act+h_{0})/PLANCK}}$&  & & & &  & & & & &  \\
  \hline
\end{tabular}
\caption{\it Cosmological parameter estimation from observationally feasible model
obtained from $\chi=-\frac{2}{3}$ branch.
 }
\label{tab4}
\end{table}
In table(\ref{tab3}) we have mentioned the various classes of {\it Jordan frame} and {\it Einstein frame}
 potentials obtained from all possible physical combinations of H and S of the superconformal transformation mentioned in equation(\ref{cvt1y}).
\begin{figure}[htb]
\centering
\subfigure[$TT$]{
    \includegraphics[width=5.5cm,height=4.7cm] {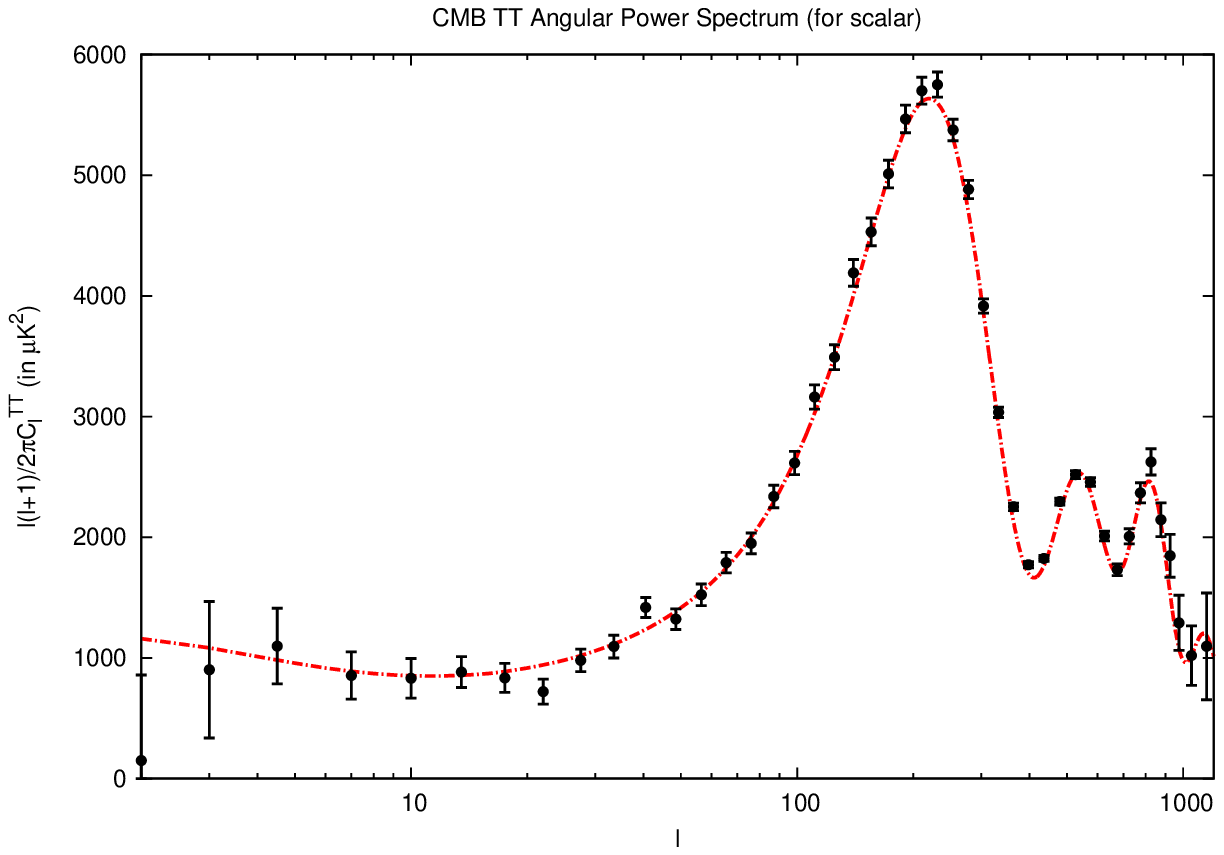}
    \label{fig:subfig4av}
}
\subfigure[$TE$]{
    \includegraphics[width=5.5cm,height=4.7cm] {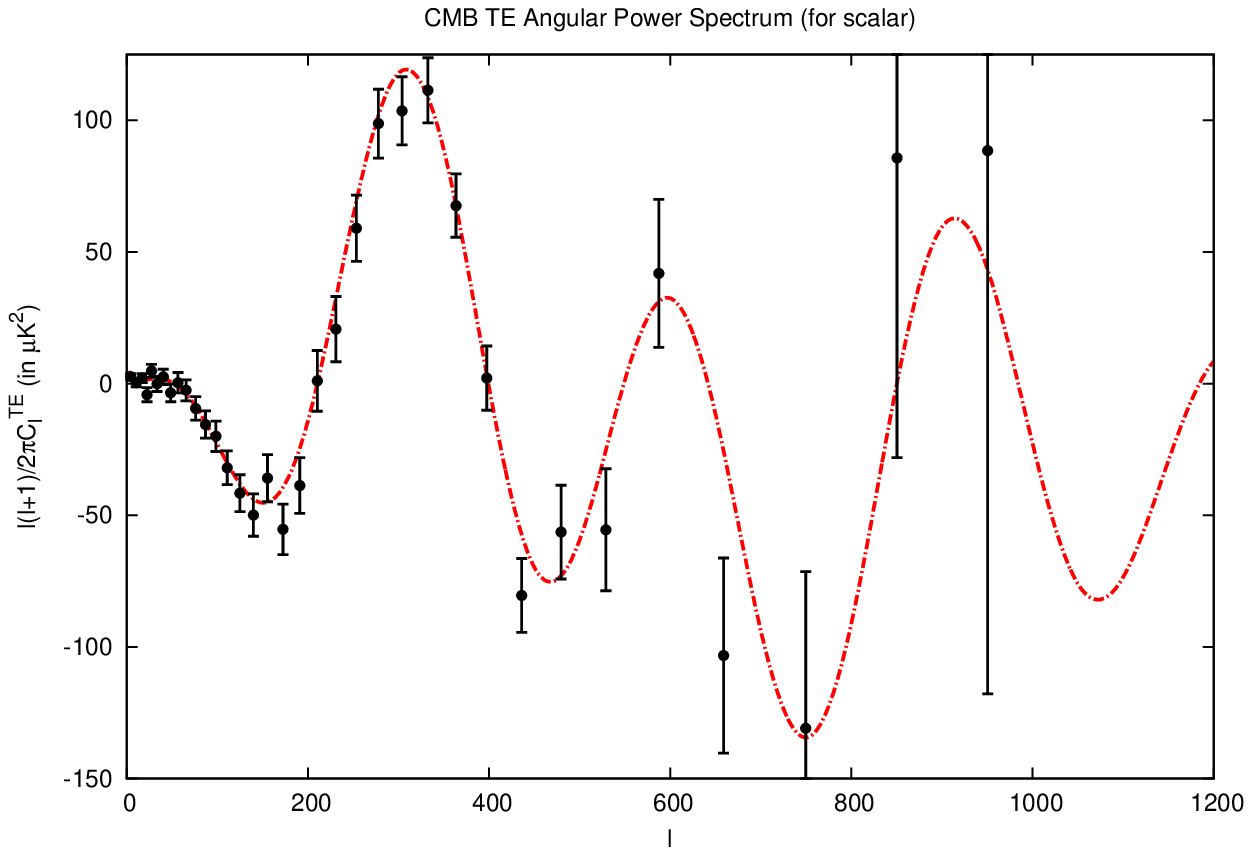}
    \label{fig:subfig5av}
}
\subfigure[$EE$]{
    \includegraphics[width=5.5cm,height=4.7cm] {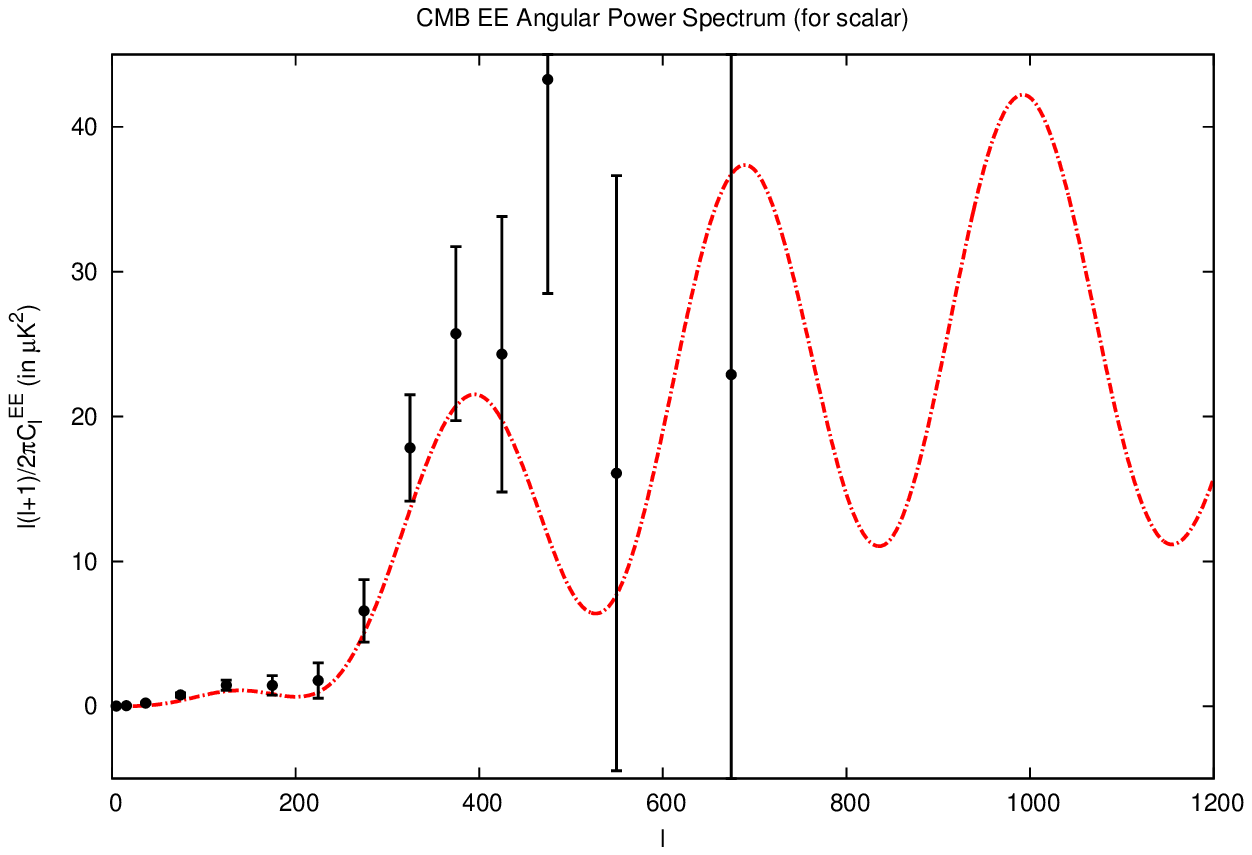}
    \label{fig:subfig6av}
}
\caption[Optional caption for list of figures]{Variation
 of CMB angular power spectra vs multipoles ($l$) for \subref{fig:subfig4av} TT, \subref{fig:subfig5av} TE
and \subref{fig:subfig6av} EE mode from $\chi=-\frac{2}{3}$ branch. The statistical error bars are obtained from WMAP9 data}
\label{fig2}
\end{figure}

In table(\ref{tab2}) and table(\ref{tab4}) we have mentioned
 all the cosmological parameters estimated from the observationally allowed potentials of $\chi=\frac{2}{3}$ and $\chi=-\frac{2}{3}$
 branch respectively. From the numerical analysis, we have explicitly shown that almost all of these proposed models confront well with latest ${\it WMAP9}$ data combined with several
complementary datasets of ${\it SPT}$, ${\it ACT}$,and ${\it h_{0}}$ observations
in the ${\it \Lambda CDM}$ background and PLANCK data set as well. Next implementing the information obtained for each model from
the cosmological code ${\it CAMB}$ we estimate dark energy density ($\Omega_{\Lambda}$), matter density ($\Omega_{m}$) and its r.m.s. fluctuation ($\sigma_{8}$) etc.
Hence we plot
the behavior of CMB angular power spectrum for ${\it TT}$, ${\it TE}$
and ${\it EE}$ polarization obtained from $\chi=\pm\frac{2}{3}$ branch as shown in figure(\ref{fig:subfig1av}-\ref{fig:subfig3av}) and
 figure(\ref{fig:subfig4av}-\ref{fig:subfig6av}) for scalar mode.

In this article our prime objective is to study the cosmological
consequences of single field inflationary potentials. For such cases the fields other than inflaton (i.e. background fields) can trigger the two phenomenological
scenarios : preheating and reheating. Further this will directly or indirectly affect the leptogenesis \cite{sayan2,sayan3,riotto,anupam5,anupam6,anupam7} and baryogenesis \cite{anupam8,anupam9} scenario depending
on the strength of the different decay channels of the inflatons into different particle constituents and the corresponding CP asymmetry of different branches.
For the precise estimation of cosmological parameters, we fix the value of all the
background fields at GUT scale ($H_{1}=H_{2}=0.9\times10^{16}~GeV$).

In this context,
all the potentials are derived from SUGRA or from its superconformal extension.
Consequently the energy scale of the potentials is around GUT scale. This directly satisfies the constraint on energy scale as
$\mu_{GUT}<\Lambda_{UV}$, where $\mu_{GUT}\sim 10^{16}GeV$ is the corresponding energy scale of SUGRA and $\Lambda_{UV}=M_{PL}$ be the UV(Ultra-Violet) cut-off theory.
Here all the {\it Yukawa} type couplings ($\lambda_{1},\lambda_{2}$) are energy scale dependent which will follow the
 {\it Renormalization Group} (RG) flow \cite{ryder,star1,star2}
via {\it Callan-Symanzik} equation. For the numerical estimation we fix the values of the
{\it Yukawa} type couplings at GUT scale in the present context. Moreover, after applying RG flow from GUT to EWSB scale all of them becomes
large ($\sim 2.065\times 10^{-3}$) imposing the experimental constraints from {\it LHC}.
It is a ray of hope for near future that proper bound on the self coupling is measurable
in the next run of the {\it LHC}. For futher details on these aspects see \cite{star1,star2} where RG flow analysis has been discussed thoroughly.
 Most importantly, the very recent Higgs mass bound observed at ${\it LHC}$
and latest observational data from ${\it WMAP9}$ and PLANCK have already ruled out the possibility of
all the proposed
inflationary potentials at the EWSB scale in the absence of any symmetry breaking non-minimal coupling. In this article by thorough numerical analysis we
explicitly show that without introducing any non-minimal coupling
 all the proposed inflationary potentials obtained from the $\chi=\pm\frac{2}{3}$ branches
are observationally favored at the GUT scale. On the other hand such running in the {\it Yukawa } type of couplings induces the possibility of {\it Primordial Black Hole} (PBH)
formation \cite{sayan1,drees1,drees2} depending on the running on the model dependent cosmological parameter $\alpha_{s}$. A very interesting fact for the inflationary model building is that
the present observation from
 PLANCK (using WMAP9 data as a prior and the complementary data set (PLANCK lensing+CMB high~${\it l}$+BAO) \cite{pl1}
has predicted $\alpha_{S}$ and  $\kappa_{S}$ to be $-0.013\pm0.009$ (although at 1.5$\sigma$) and $0.020^{+0.016}_{-0.015}$ respectively.
 Additionally for both $\chi=\pm\frac{2}{3}$
 branches tensor to scalar ratio ($r$) are within the observational upper
 bound of PLANCK.


\subsection{\bf Models with $\chi\neq\pm\frac{2}{3}$}
In this context the symmetry breaking parameter $\chi$ is connected with the non-minimal coupling $\xi$ present
as $\frac{\xi}{2}\phi^{2}R$ in the action. To explore more features from this sector we consider two physical situations
given by:
\be\begin{array}{llll}\label{ki}
    \displaystyle \chi-\frac{2}{3}=4\xi_{1}
\end{array}\ee
\be\begin{array}{llll}\label{kq} \displaystyle \chi +\frac{2}{3}=4\xi_{2}
   \end{array}\ee

where $\xi_{1}$ and $\xi_{2}$ are the two non-minimal couplings approaching from $\frac{2}{3}$ and $-\frac{2}{3}$ respectively.

From equation(\ref{ki}) and equation(\ref{kq}) the superconformal factors can be expressed as:
\be\begin{array}{llll}\label{hj4}
    \displaystyle \Omega^{2}_{1}(H,\bar{H},S,\bar{S})=1+\frac{1}{2}\left(\xi_{1}+\frac{1}{3}\right)\left[(H-\bar{H})^{2}+(S-\bar{S})^{2}\right]+\frac{\xi_{1}}{2}\left[(H+\bar{H})^{2}+(S+\bar{S})^{2}\right]
   \end{array}\ee
\be\begin{array}{llll}\label{hj4za}
    \displaystyle \Omega^{2}_{2}(H,\bar{H},S,\bar{S})=1+\frac{\xi_{2}}{2}\left[(H-\bar{H})^{2}+(S-\bar{S})^{2}\right]+\frac{1}{2}\left(\xi_{2}-\frac{1}{3}\right)\left[(H+\bar{H})^{2}+(S+\bar{S})^{2}\right]
   \end{array}\ee

In table(\ref{tab5}) and table(\ref{tab7}) we mention all types of inflationary potentials in {\it Jordan frame} and
{\it Einstein frame} as obtained from the
 two possible physical branches of the superconformal
transformations mentioned in equation(\ref{hj4}) and equation(\ref{hj4za}) respectively.

\begin{table}[htb]\small\addtolength{\tabcolsep}{-1pt}
   \small\begin{tabular}{|c|c|c|c|c|}
   \hline ${\bf Class~of~models}$  &{\bf $\Omega^{2}_{1}$ }
&  ${\bf {\cal W}}$ &{\bf $V_{J}$}
&  {\bf $V_{E}$}  \\
\hline H~real,S=0 & $\left(1+\xi_{1} H^{2}_{1}\right)$ & -$\lambda_{1} S\left(H\bar{H}-\frac{v^{2}_{1}}{2}\right)$
 &$\frac{\lambda^{2}_{1}}{4}\left(H^{2}_{1}-v^{2}_{1}\right)^{2}$ & $\frac{\frac{\lambda^{2}_{1}}{4}\left(H^{2}_{1}-v^{2}_{1}\right)^{2}}
{\left(1+\xi_{1} H^{2}_{1}\right)^{2}}$\\
   \hline
 H=0,S~real  &$\left(1+\xi_{1} S^{2}_{1}\right)$ & -$\lambda_{2} H\left(S\bar{S}-\frac{v^{2}_{2}}{2}\right)$ &$\frac{\lambda^{2}_{2}}{4}\left(S^{2}_{1}-v^{2}_{2}\right)^{2}$
 & $\frac{\frac{\lambda^{2}_{2}}{4}\left(S^{2}_{1}-v^{2}_{2}\right)^{2}}{\left(1+\xi_{1} S^{2}_{1}\right)^{2}}$ \\
    \hline
 H~complex,S=0  & $1-\left(\xi_{1}+\frac{1}{3}\right)H^{2}_{2}$ & -$\lambda_{1} S\left(H\bar{H}-\frac{v^{2}_{1}}{2}\right)$
&$\frac{\lambda^{2}_{1}}{4}\left(H^{2}_{1}+H^{2}_{2}-v^{2}_{1}\right)^{2}$ &$\frac{\frac{\lambda^{2}_{1}}{4}\left(H^{2}_{1}+H^{2}_{2}
-v^{2}_{1}\right)^{2}}{\left[1-\left(\xi_{1}+\frac{1}{3}\right)H^{2}_{2}+\xi_{1}H^{2}_{1}\right]^{2}}$ \\
& $+\xi_{1}H^{2}_{1}$ & & & \\
 \hline H=0,S~complex& $1-\left(\xi_{1}+\frac{1}{3}\right)S^{2}_{2}$&-$\lambda_{2} H\left(S\bar{S}-\frac{v^{2}_{2}}{2}\right)$ & $\frac{\lambda^{2}_{2}}{4}\left(S^{2}_{1}+S^{2}_{2}-v^{2}_{2}\right)^{2}$
&$\frac{\frac{\lambda^{2}_{2}}{4}\left(S^{2}_{1}+S^{2}_{2}-v^{2}_{2}\right)^{2}}{\left[1-\left(\xi_{1}+\frac{1}{3}\right)S^{2}_{2}+\xi_{1}S^{2}_{1}\right]^{2}}$ \\
& $+\xi_{1}S^{2}_{1}$ & & & \\
   \hline
\end{tabular}
  \caption{\it Jordan frame and Einstein frame potentials
obtained from $\chi-\frac{2}{3}=4\xi_{1}$ branch.}
  \label{tab5}
  \end{table}
\begin{table}[htb]\small\addtolength{\tabcolsep}{-1pt}
\small\begin{tabular}{|l|l|l|l|l|l|l|l|l|l|l|l|l|}
  \hline
  {\bf Potential}& {\bf Confronts } & {\bf Coup} & {\bf $\xi_{1}$} & { \bf $P_{R}$}&{\bf $n_{s}$}& {\bf $\alpha_{s}$}& {\bf $r$ }&
 {\bf $\Omega_{\Lambda}$}&{\bf $\Omega_{m}$}&{\bf $\sigma_{8}$}&{\bf $\eta_{Rec}$} &{\bf $\eta_{0}$}\\
 &{\bf with}&{\bf-lings} & &{\bf ($\times 10^{-9}$)} & &{\bf ($\times 10^{-4}$)} &   & &  & &{\bf Mpc} &{\bf Mpc} \\
& &{\bf($\times 10^{-6}$)} & &  & & & & & & & &    \\
\hline
   $\frac{\frac{\lambda^{2}_{1}}{4}\left({\bf{H^{2}_{1}}}-v^{2}_{1}\right)^{2}}{\left(1+\xi_{1}
 {\bf{H^{2}_{1}}}\right)^{2}}$&${\bf{ \Lambda CDM(WMAP9+spt}}$ & $5.167$ & 0.1 & $2.330$ & $0.961$&$-11.752 $&$0.015$
& $0.684$&$0.316$ &$0.821 $ & $280.38$ & $14184.8$  \\
&${\bf{ +act+h_{0})/PLANCK}}$ & & & & & &  & & & & & \\
   \hline
  $\frac{\frac{\lambda^{2}_{1}}{4}\left({\bf{H^{2}_{1}}}+H^{2}_{2}
-v^{2}_{1}\right)^{2}}{\left[1-\left(\xi_{1}+\frac{1}{3}\right)H^{2}_{2}+\xi_{1}{\bf{H^{2}_{1}}}\right]^{2}}$ &${\bf{ \Lambda CDM(WMAP9+spt}}$
& $6.789$ & 0.1 & $2.310$ & $0.960$ & $-9.94$ & $0.013$  & $0.684$ & $0.316$ &$0.816$  & $280.38$ & $14184.8$  \\
 &${\bf{ +act+h_{0})/PLANCK}}$& & & & & &  & & & & & \\
   \hline
  $\frac{\frac{\lambda^{2}_{2}}{4}\left(H^{2}_{1}+{\bf{H^{2}_{2}}}
-v^{2}_{2}\right)^{2}}{\left[1-\left(\xi_{1}+\frac{1}{3}\right){\bf{H^{2}_{2}}}+\xi_{1}H^{2}_{1}\right]^{2}}$& ${\bf{ \Lambda CDM(WMAP9+spt}}$
 & $5.818$  & $-0.5$ & $2.318$ & $0.962$ & $-8.801$ & $0.011$  &
 $0.684$ & $0.316$ & $0.821$ & $280.38$ & $14184.8$ \\
 & ${\bf{ +act+h_{0})/PLANCK}}$  &  &  & & & &  & &  & & &   \\

  \hline
\end{tabular}
\caption{\it Cosmological parameter estimation from observationally allowed models
obtained from $\chi-\frac{2}{3}=4\xi_{1}$ branch.}
\label{tab6}
\end{table}
\begin{figure}[htb]
\centering
\subfigure[$TT$]{
    \includegraphics[width=5.5cm,height=4.7cm] {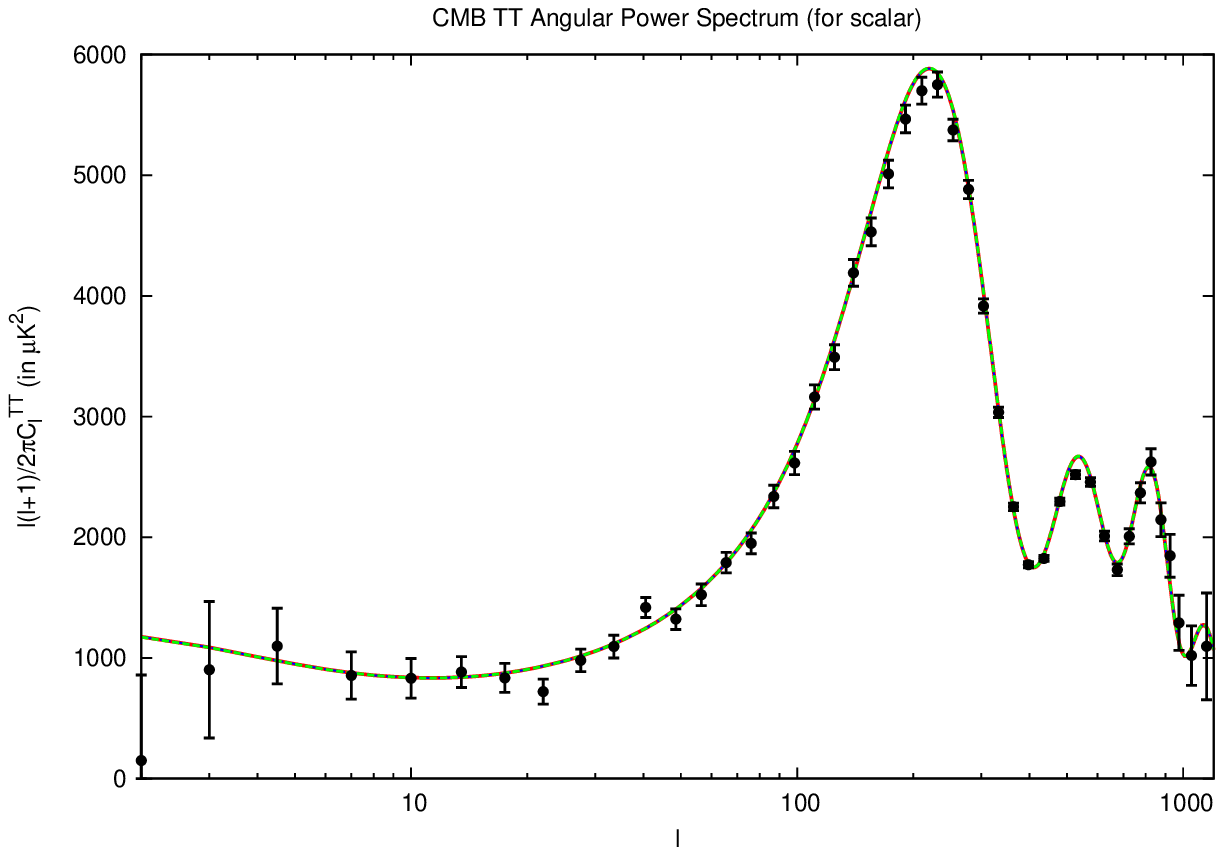}
    \label{fig:subfig7av}
}
\subfigure[$TE$]{
    \includegraphics[width=5.5cm,height=4.7cm] {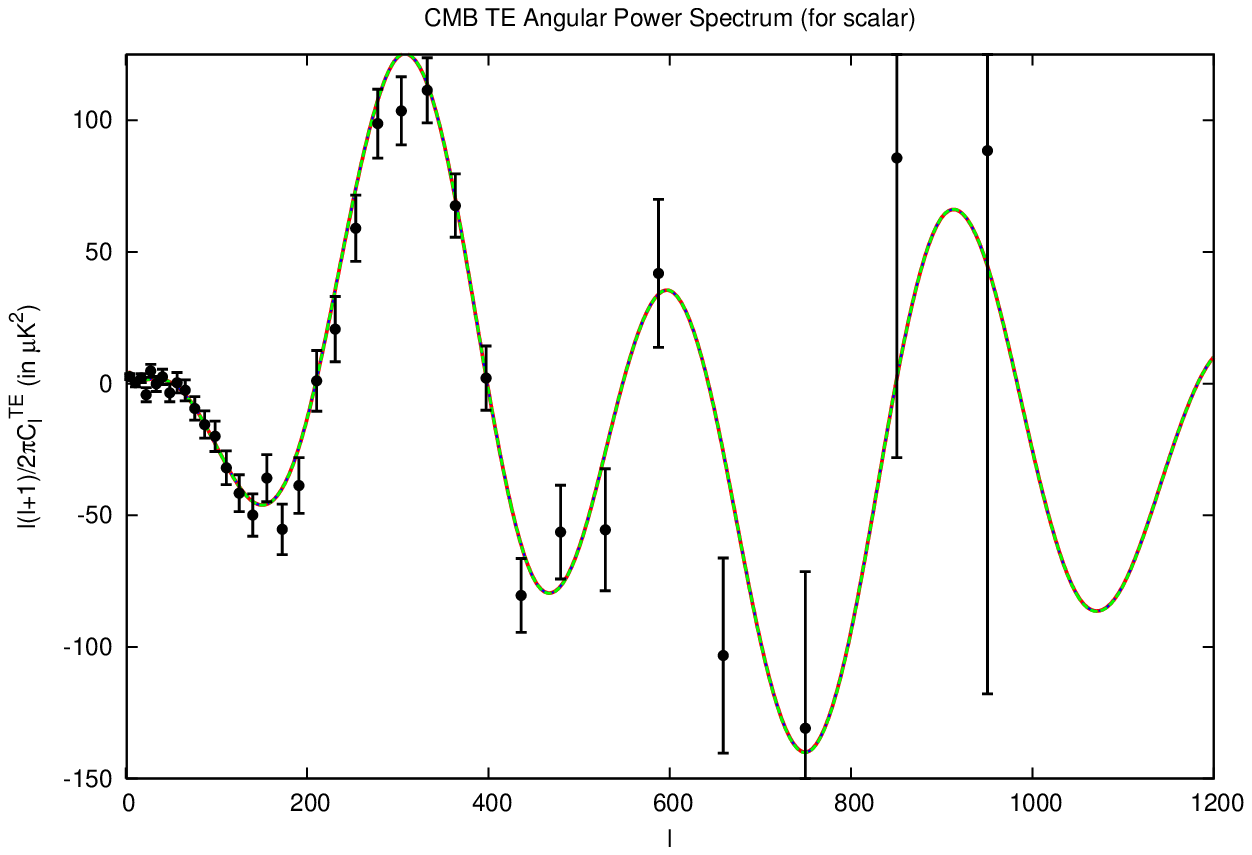}
    \label{fig:subfig8av}
}
\subfigure[$EE$]{
    \includegraphics[width=5.5cm,height=4.7cm] {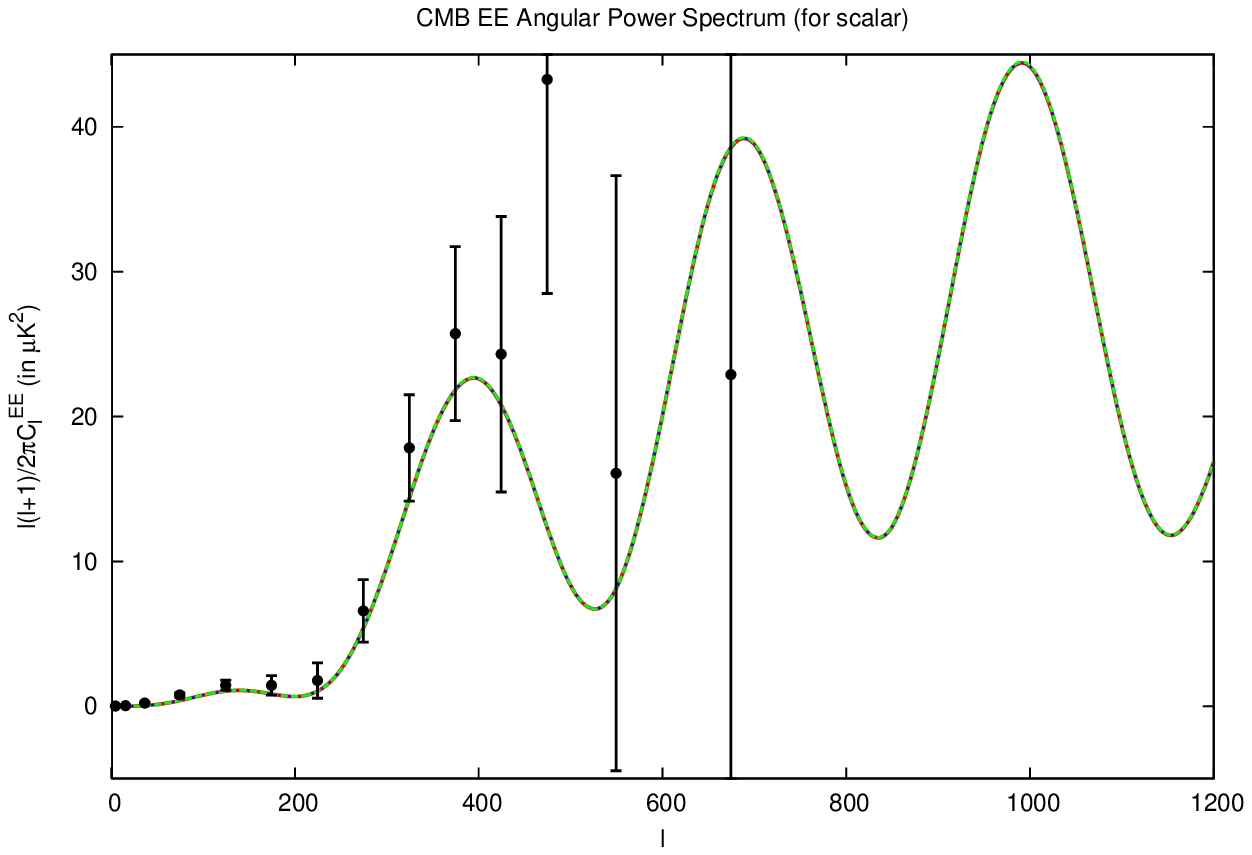}
    \label{fig:subfig9av}
}
\caption[Optional caption for list of figures]{Variation
 of CMB angular power spectra vs multipoles ($l$) \subref{fig:subfig7av} TT, \subref{fig:subfig8av} TE and
\subref{fig:subfig9av}EE mode from $\chi-\frac{2}{3}=4\xi_{1}$ branch. The statistical error bars are obtained from WMAP9 data.}
\label{fig3}
\end{figure}

\begin{table}[htb]\small\addtolength{\tabcolsep}{-1pt}
   \small\begin{tabular}{|c|c|c|c|c|}
   \hline ${\bf Class~of~models}$  &{\bf $\Omega^{2}_{2}$ }
&  ${\bf {\cal W}}$ &{\bf $V_{J}$}
&  {\bf $V_{E}$}  \\
\hline H~real,S=0 & $\left[1+\left(\xi_{2}-\frac{1}{3}\right) H^{2}_{1}\right]$ & -$\lambda_{1} S\left(H\bar{H}-\frac{v^{2}_{1}}{2}\right)$
 &$\frac{\lambda^{2}_{1}}{4}\left(H^{2}_{1}-v^{2}_{1}\right)^{2}$ & $\frac{\frac{\lambda^{2}_{1}}{4}\left(H^{2}_{1}-v^{2}_{1}\right)^{2}}
{\left[1+\left(\xi_{2}-\frac{1}{3}\right) H^{2}_{1}\right]^{2}}$\\
   \hline
 H=0,S~real  &$\left[1+\left(\xi_{2}-\frac{1}{3}\right) S^{2}_{1}\right]$ & -$\lambda_{2} H\left(S\bar{S}-\frac{v^{2}_{2}}{2}\right)$ &$\frac{\lambda^{2}_{2}}{4}\left(S^{2}_{1}-v^{2}_{2}\right)^{2}$
 & $\frac{\frac{\lambda^{2}_{2}}{4}\left(S^{2}_{1}-v^{2}_{2}\right)^{2}}{\left[1+\left(\xi_{2}-\frac{1}{3}\right) S^{2}_{1}\right]^{2}}$ \\
    \hline
 H~complex,S=0  & $1+\left(\xi_{2}-\frac{1}{3}\right)H^{2}_{1}$ & -$\lambda_{1} S\left(H\bar{H}-\frac{v^{2}_{1}}{2}\right)$
&$\frac{\lambda^{2}_{1}}{4}\left(H^{2}_{1}+H^{2}_{2}-v^{2}_{1}\right)^{2}$ &$\frac{\frac{\lambda^{2}_{1}}{4}\left(H^{2}_{1}+H^{2}_{2}
-v^{2}_{1}\right)^{2}}{\left[1+\left(\xi_{2}-\frac{1}{3}\right)H^{2}_{1}-\xi_{2}H^{2}_{2}\right]^{2}}$ \\
& $-\xi_{2}H^{2}_{2}$ & & & \\
 \hline H=0,S~complex& $1+\left(\xi_{2}-\frac{1}{3}\right)S^{2}_{1}$&-$\lambda_{2} H\left(S\bar{S}-\frac{v^{2}_{2}}{2}\right)$ & $\frac{\lambda^{2}_{2}}{4}\left(S^{2}_{1}+S^{2}_{2}-v^{2}_{2}\right)^{2}$
&$\frac{\frac{\lambda^{2}_{2}}{4}\left(S^{2}_{1}+S^{2}_{2}-v^{2}_{2}\right)^{2}}{\left[1+\left(\xi_{2}-\frac{1}{3}\right)S^{2}_{1}-\xi_{2}S^{2}_{2}\right]^{2}}$ \\
& $-\xi_{2}S^{2}_{2}$ & & & \\
   \hline
\end{tabular}
  \caption{\it Jordan frame and Einstein frame potentials
obtained from $\chi+\frac{2}{3}=4\xi_{2}$ branch. }
\label{tab7}
\end{table}


\begin{table}[htb]\tiny\addtolength{\tabcolsep}{-1pt}
\tiny\begin{tabular}{|l|l|l|l|l|l|l|l|l|l|l|l|l|l|l|}
  \hline
  {\bf Potential} & {\bf Confronts }& {\bf Coup} & {\bf $\xi_{2}$} & { \bf $P_{R}$}&{\bf $n_{s}$}& {\bf $\alpha_{s}$}& {\bf $r$ }&
 {\bf $\Omega_{\Lambda}$}&{\bf $\Omega_{m}$}&{\bf $\sigma_{8}$} &{\bf $\eta_{Rec}$} &{\bf $\eta_{0}$}\\
&{\bf with} &{\bf-lings} & &{\bf ($\times 10^{-9}$)} & &{\bf ($\times 10^{-4}$)} &  &  &  & &{\bf Mpc} &{\bf Mpc} \\
& &{\bf($\times 10^{-6}$)} & &  & & & & & & & &   \\
\hline
  $\frac{\frac{\lambda^{2}_{1}}{4}\left({\bf{H^{2}_{1}}}-v^{2}_{1}\right)^{2}}
{\left[1+(\xi_{2}-\frac{1}{3} ){\bf H^{2}_{1}}\right]^{2}}$&${\bf {\Lambda CDM(WMAP9+spt}}$  & $9.254$ & 0.5 & $2.370$ & $0.960$&$-10.006 $
&$0.018$  & $0.684$ & $0.316$ & $0.826 $ &  $280.38$ & $14184.8$  \\
 &${\bf{ +act+h_{0})/PLANCK}}$& & & & & & & & & & &  \\
  \hline
  $\frac{\frac{\lambda^{2}_{1}}{4}\left({\bf H^{2}_{1}}+H^{2}_{2}
-v^{2}_{1}\right)^{2}}{\left[1+\left(\xi_{2}-\frac{1}{3}\right){\bf H^{2}_{1}}-\xi_{2}H^{2}_{2}\right]^{2}}$  & ${\bf{\Lambda CDM(WMAP9+spt}}$
 & $7.152$ & 0.5 & $2.32$ & $0.961$ & $-9.712$ & $0.011$  & $0.684$ & $0.316$ & $0.818 $ & $280.38$ & $14184.8$  \\
&${\bf {+act+h_{0})/PLANCK}}$ & & & & & &  & & & & &  \\
 
   \hline
 
  $\frac{\frac{\lambda^{2}_{1}}{4}\left(H^{2}_{1}+{\bf H^{2}_{2}}
-v^{2}_{1}\right)^{2}}{\left[1+\left(\xi_{2}-\frac{1}{3}\right)H^{2}_{1}-\xi_{2}{\bf H^{2}_{2}}\right]^{2}}$ & ${\bf{ \Lambda CDM(WMAP9+spt}}$ & $5.184$  &
$-0.1$ & $2.340$ & $0.961$ & $-9.365$ & $0.015$ & $0.684$ & $0.316$ & $0.822$ &  $280.38$ & $14184.8.2$ \\
  & ${\bf{ +act+h_{0})/PLANCK}}$ &  &  & & & &  & &  & & &    \\
  
  \hline
\end{tabular}
\caption{\it Cosmological parameter estimation for observationally favored models
obtained from $\chi+\frac{2}{3}=4\xi_{2}$ branch.}
\label{tab8}
\end{table}
\begin{figure}[htb]
\centering
\subfigure[$TT$]{
    \includegraphics[width=5.5cm,height=4.7cm] {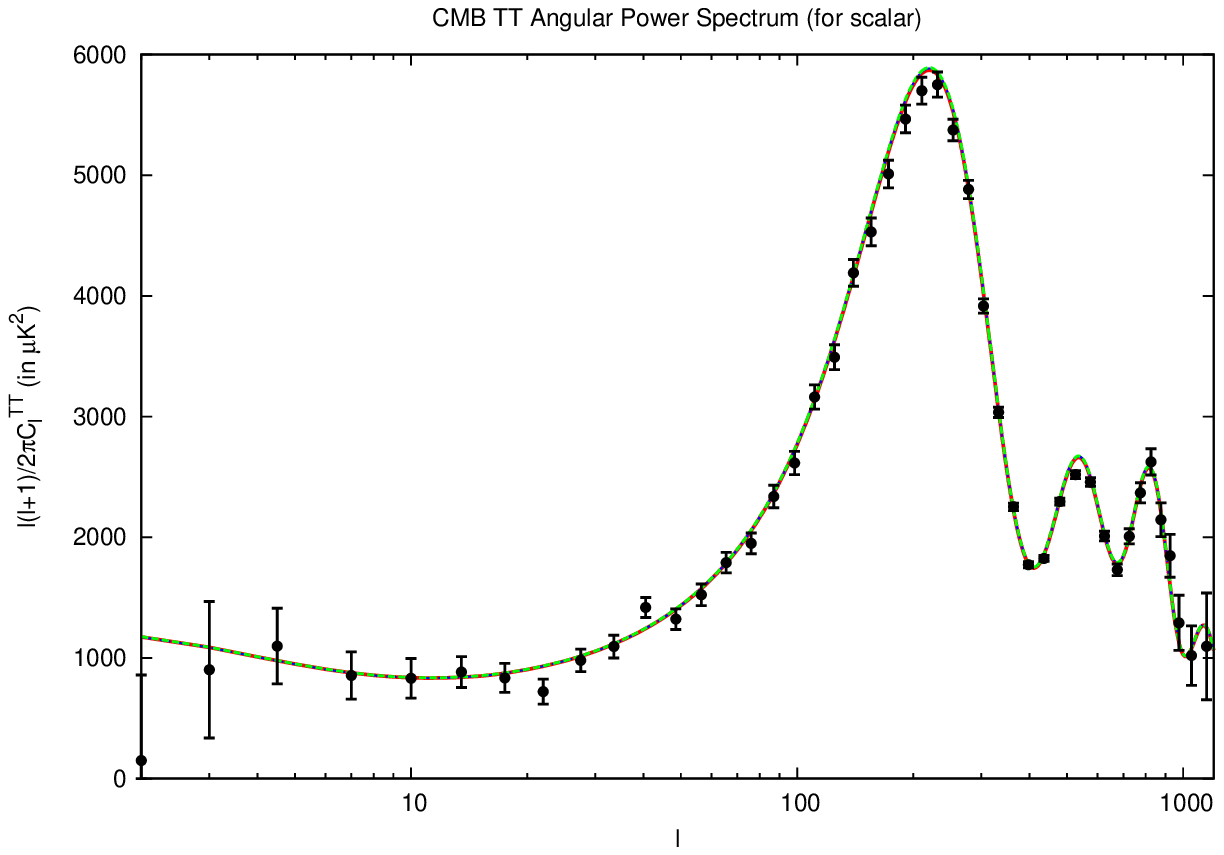}
    \label{fig:subfig10av}
}
\subfigure[$TE$]{
    \includegraphics[width=5.5cm,height=4.7cm] {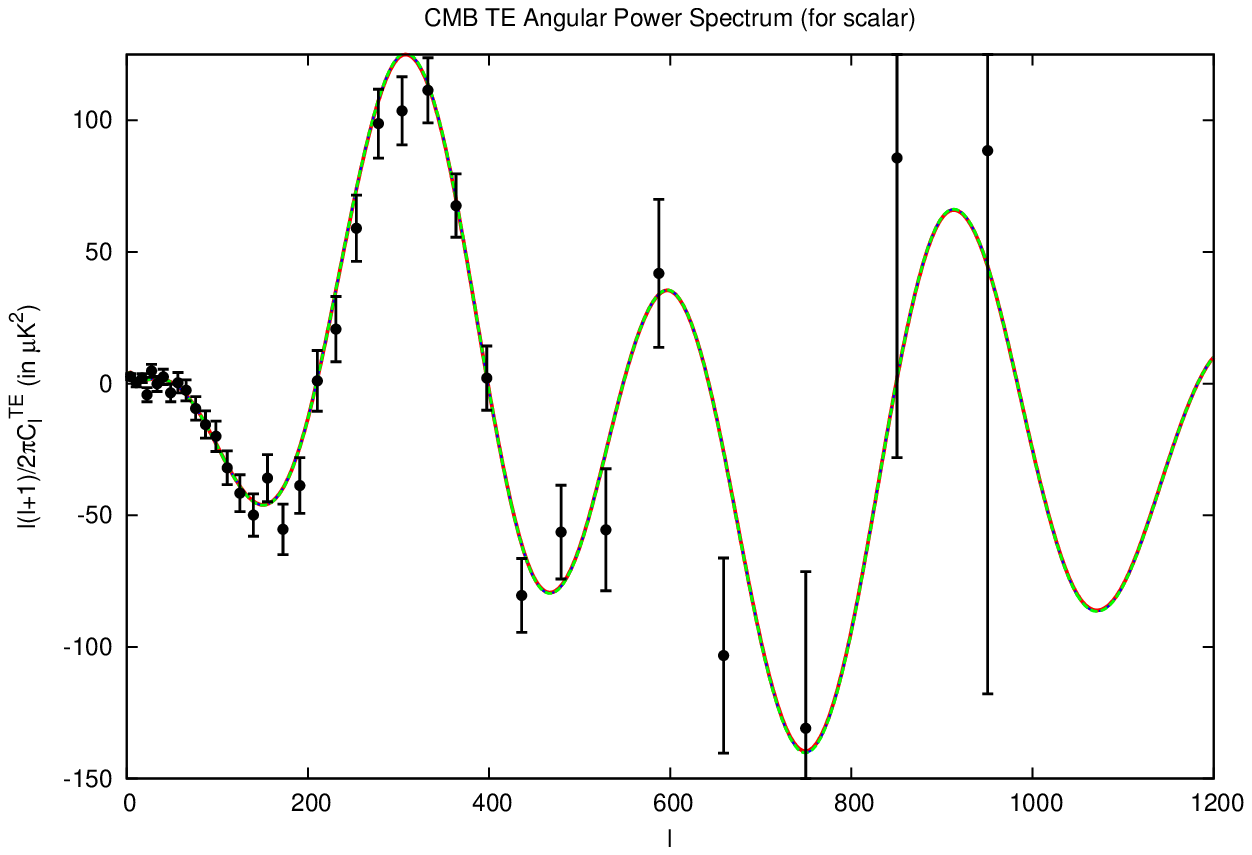}
    \label{fig:subfig11av}
}
\subfigure[$EE$]{
    \includegraphics[width=5.5cm,height=4.7cm] {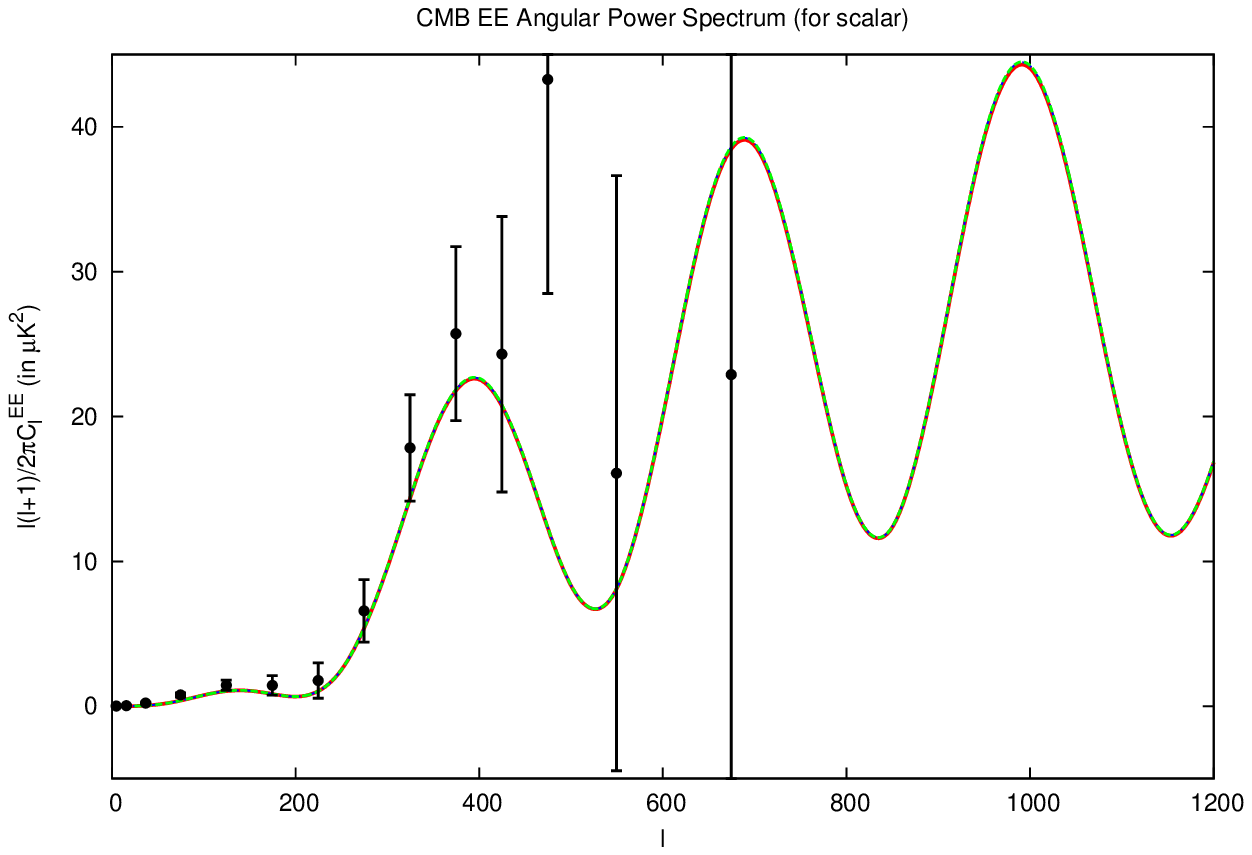}
    \label{fig:subfig12av}
}
\caption[Optional caption for list of figures]{Variation
 of CMB angular power spectra vs multipoles ($l$) for \subref{fig:subfig10av} TT, \subref{fig:subfig11av} TE
\subref{fig:subfig12av}EE mode from $\chi+\frac{2}{3}=4\xi_{2}$ branch. The statistical error bars are obtained from WMAP9 data.}
\label{fig3a}
\end{figure}

\begin{figure}[htb]
{\centerline{\includegraphics[width=13cm, height=8cm] {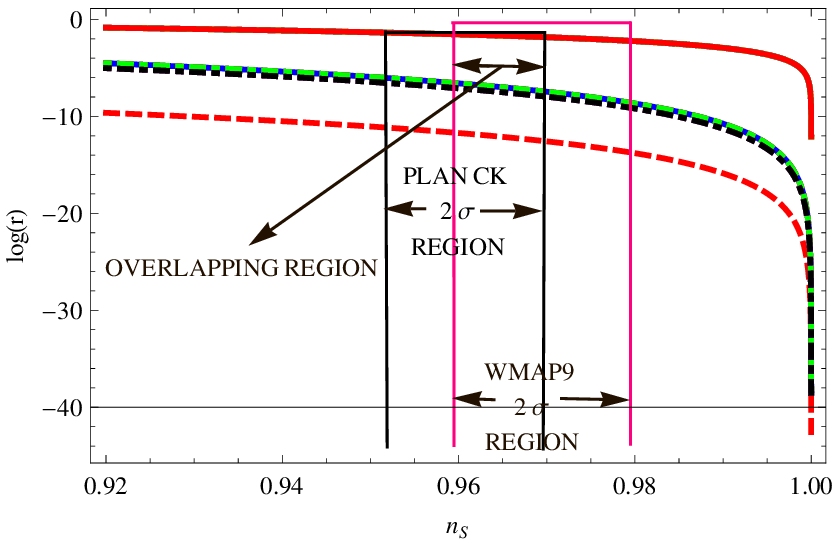}}}
\caption{Variation of tensor to scalar ratio ($r$) vs scalar spectral index ($n_{S}$) for the family of Higgs potentials for
different numerical values of non-minimal coupling $\xi$. The value of the non-minimal coupling increases as we go down towards the plot. This also shows
 $\chi\pm\frac{2}{3}=4\xi$ branches are more observationally favored compared to the $\chi=\pm\frac{2}{3}$ branches.
  } \label{figns}
\end{figure}

Next we have mentioned all the cosmological parameters estimated from $\chi\neq\frac{2}{3}$ ($\chi-\frac{2}{3}=4\xi_{1}$ and $\chi+\frac{2}{3}=4\xi_{2}$) branches in table(\ref{tab6}) and table(\ref{tab8}).
This clearly shows non-minimal coupling ($\xi_{1},\xi_{2}$) dependent models confront with latest data.
We have also shown that if we allow the above mentioned non-minimal couplings along with very recent {\it LHC} Higgs mass bound and
latest observational constraints, then
almost all of the proposed
inflationary potentials are favored starting from EWSB to GUT scale depending on the RG flow in {\it Yukawa} type coupling.
Throughout the numerical analysis we have allowed both the signatures of the non-minimal coupling. We also avoided specific values of the non-minimal couplings for which
divergences are appearing in the proposed potentials. During the analysis we have observed that only for $(\xi_{1},\xi_{2})>0$ the first two models appearing in
table(\ref{tab6}) and table(\ref{tab8}) are in good agreement with
latest observation. On the contrary for $(\xi_{1},\xi_{2})<0$ only the third model fairs well with {\it WMAP9} and PLANCK data set.
Moreover, for the numerical estimations we consider only those values of the non-minimal couplings for which the proposed models
are free from any poles.
The behavior of tensor to scalar ratio ($r$) with respect to the scalar spectral index ($n_{S}$) for all class
of proposed models of inflation are depicted in figure(\ref{figns}).

\section{Summary and outlook}

In this article we have proposed a class
of supergravity motivated models to implement Higgs
inflation, where the Higgs field is non-minimally coupled
to gravity sector via symmetry breaking coupling ($\chi$).
We have followed the analysis by making use of superconformal techniques
in the K$\ddot{a}$hler manifold. Using such tools we have introduced a phenomenological
K$\ddot{a}$hler potential which preserves shift symmetry for two minimal coupling
$\chi=\pm\frac{2}{3}$ with gravity.
This results in various classes of inflationary models which are made up of shift symmetry protected flat directions.
 We have elaborately discussed the consequences of superconformal techniques in the two
 preferred frame of references namely, Jordan and Einstein frames.
Then we have explored the features of non-minimal coupling ($\xi_{1},\xi_{2}$)
connected with shift symmetry breaking branch $\chi\neq\frac{2}{3}$ in the context of Higgs inflation.
Hence we have studied inflation from these proposed models
by estimating the observable parameters which originates from primordial quantum fluctuation for scalar and tensor modes.
We have further confronted our results with WMAP9 and various complementary dataset ($SPT,ACT,h_{0}$) by
using CAMB and as well as independently with PLANCK data set. Further we have compared the behavior of theoretical CMB polarization power spectra for $TT,TE$ and $EE$ mode obtained
from all of these proposed models with observational power spectra.
We have also commented on the allowed range for non-minimal couplings ($\xi_{1},\xi_{2}$)
and phenomenological {\it Yukawa} type of couplings which are very crucial inputs in the context of inflationary model building.
This, collectively, provides an exhaustive study of the class of Higgs inflation from K$\ddot{a}$hler potential and consequently, their pros and cons.

An interesting open issue in this context is to study the role of Hiesenberg symmetry \cite{stefan1,stefan2,olive}.
Other open issues is to study primordial black hole formation and its cosmological consequences from
the running of the spectral index ($\alpha_{S}$) and its running ($\kappa_{S}$) as the very recent
{\it PLANCK} data gives an estimation of the above mentioned indexes at $1.5\sigma$ \cite{pl1}. Moreover, the phenomenological consequences of
all of these proposed models via reheating and leptogenesis are also a promising issue for future study.


\section*{Acknowledgments}

SC thanks Council of Scientific and
Industrial Research, India for financial support through Senior
Research Fellowship (Grant No. 09/093(0132)/2010). SC also thanks Soumya Sadhukhan for useful discussions.
TC thanks Council of Scientific and
Industrial Research, India for financial support through Senior
Research Fellowship (Grant No. 09/093(0134)/2010).




\end{document}